%% file: manuscript/main.tex
\documentclass[pdflatex,iicol,sn-mathphys-num]{sn-jnl}

\usepackage{graphicx}%
\usepackage{multirow}%
\usepackage{amsmath,amssymb,amsfonts}%
\usepackage{amsthm}%
\usepackage{mathrsfs}%
\usepackage[title]{appendix}%
\usepackage{xcolor}%
\usepackage{textcomp}%
\usepackage{manyfoot}%
\usepackage{booktabs}%
\usepackage{algorithm}%
\usepackage{algorithmicx}%
\usepackage{algpseudocode}%
\usepackage{listings}%
\usepackage{cleveref}%
\usepackage{makecell}
\usepackage{placeins}
\usepackage{xurl}
\AtBeginDocument{}
\AtBeginDocument{%
  }

\theoremstyle{thmstyleone}%
\theoremstyle{thmstyletwo}%

\theoremstyle{thmstylethree}%

\begin{document}

\title[Horizon-Aware EEP for Tokamak Disruption Alarms]{Horizon-Aware Early Event Prediction for Tokamak Disruption Alarms}


\author*[1,2]{Takeshi Koshizuka}\email{takeshi.koshizuka@riken.jp}
\author[1,2]{Takaharu Yaguchi}\email{yaguchi@imi.kyushu-u.ac.jp}

\affil*[1]{RIKEN AIP, Japan}
\affil[2]{Kyushu University IMI, Fukuoka, Japan}


\abstract{
Disruption predictors must warn the plasma control system early enough for the intended response, and the required warning time differs between mitigation and avoidance. We view a warning system as a predictor that outputs a disruption risk and an alarm policy that acts on it, and ask which risk the predictor should learn. Binary classifiers do not specify when a disruption occurs, and survival analysis models the full distribution of the time to disruption, whereas an alarm depends only on the risk within a finite horizon. We introduce two horizon-focused targets into tokamak disruption prediction: a horizon-truncated survival objective and fixed-horizon early event prediction (EEP) with temporal label smoothing (TLS). We compare them with binary classification and survival analysis on the DIII-D, Alcator C-Mod, and EAST databases of DisruptionBench, evaluating the risk by its threshold-free ranking at a 40 ms deadline and the alarm policy by thresholds calibrated on validation data. On every device, TLS outperforms the deep survival models in both ranking and alarm performance; TLS performs best on DIII-D and EAST, and a random forest trained on a hard horizon target performs best on C-Mod. TLS also improves the calibrated alarm performance over hard targets placed at either end of its smoothing range on all devices, isolating the benefit of label smoothing. Horizon truncation alone did not consistently improve survival analysis. Alarm performance remains limited on C-Mod and DIII-D and varies strongly across training seeds, and ablations show that the target should extend to the required warning time.
}




\maketitle

\input{manuscript/introduction}
\input{manuscript/related_work}
\input{manuscript/method}
\input{manuscript/experiments}
\input{manuscript/conclusion}

\backmatter

\bmhead{Acknowledgements}

This work was supported by JST Moonshot R\&D Grant Number JPMJMS24A3,
JST CREST Grant Number JPMJCR24Q5, and JST ASPIRE Grant Number
JPMJAP2329, Japan, and by the Horizon Europe Marie Sk{\l}odowska-Curie
Actions Staff Exchanges project 101131557 (REMODEL).

\input{manuscript/appendix}


\bibliography{bibliography/references}

\end{document}

%% file: manuscript/introduction.tex
\section{Introduction}
\label{sec:introduction}

Disruptions remain one of the main threats to the safe operation of
future tokamaks. A disruption can impose severe thermal and
electromagnetic loads on plasma-facing and structural components, and
the plasma control system (PCS) must therefore be warned early enough to
select and execute a response \citep{rea2019realtime,keith2024riskaware}.
The required warning time depends on that response: triggering the
disruption mitigation system requires only a short lead time, whereas
avoidance actions such as a controlled ramp-down require substantially
more \citep{keith2024riskaware}.

A data-driven disruption warning system can be viewed as two modules.
A predictor maps the measurements available at each time to a
disruption risk, a quantity with a defined meaning, and an alarm policy
decides from this risk when to raise an alarm, for example by
thresholding it with hysteresis
\citep{rea2019realtime,spangher2025disruptionbench}. An alarm policy can
be designed and calibrated for an intended response only if the risk is
defined in the terms that this response requires. The central question
is therefore which quantity the predictor should be trained to output.

Most data-driven disruption predictors are binary classifiers trained
to recognize a pre-disruptive state, defined by the outcome of the
discharge or by a class time before the disruption
\citep{rea2019realtime,katesharbeck2019predicting,zhu2021hybrid,spangher2025disruptionbench}.
Their output indicates whether the plasma appears to be heading towards
a disruption, but not when the disruption will occur. The class time is
usually chosen to separate stable and pre-disruptive phases rather than
from the lead time of a control action, and a single class time does
not match the precursor timescales of all disruption mechanisms
\citep{keith2024riskaware}. To select and time a response, the PCS
instead needs the risk that a disruption occurs within the time required
by the intended action. Survival analysis models this risk directly: by
modeling the conditional distribution of the time to disruption while
accounting for right censoring \citep{kalbfleisch2002statistical}, it
provides the disruption risk within any horizon from a single model
\citep{tinguely2019survival,keith2024riskaware}.
\citet{keith2024riskaware} formulated risk-aware disruption prediction in
this framework and applied deep survival models, including deep Cox
proportional hazards (DCPH) \citep{faraggi1995neural} and Deep Survival
Machines (DSM) \citep{nagpal2021dsm}, to Alcator C-Mod and DIII-D.

Whether such a risk can be learned reliably from data is a separate
question. A survival model fits the distribution of the residual time
over its full range. In long discharges, most landmarks lie far from the
disruption, where precursor signals are weak, so much of the likelihood
concerns event times well beyond the horizon relevant to an alarm,
although the alarm decision depends only on the risk within that
horizon. One remedy keeps the survival model but truncates its objective
at the horizon, as in the horizon-truncated dynamic survival analysis of
\citet{yeche2024dsa}, which we refer to as survTLS. A further step gives
up modeling the time to disruption and learns only the risk within the
horizon. This fixed-horizon early event prediction (EEP), developed for
clinical early-warning systems, estimates at every time step whether the
event will occur within a prescribed horizon
\citep{tomasev2019aki,hyland2020circulatory,yeche2023tls}. Its hard
target, however, switches from negative to positive at a single residual
time, although precursor signals do not change abruptly there, and only
the last part of each disruptive discharge is positive, so positive
examples are scarce. Temporal label smoothing (TLS) \citep{yeche2023tls}
addresses both issues by replacing the abrupt boundary with a target
that decreases smoothly with the time to the event. To our knowledge, neither formulation has been examined for
tokamak disruption prediction.

Which of these risks best serves an alarm policy has not been
established for disruption prediction. In this work, we therefore
compare predictors trained on binary classification (BC),
full-distribution survival analysis (SA), horizon-truncated survival
analysis, and fixed-horizon EEP with hard and smoothed targets, on data
from three tokamaks.

The two-module view also determines how the predictors should be
evaluated. A risk is useful only if it separates discharges that will
disrupt from those that will not early enough for the intended action,
and this property does not depend on the threshold that a particular
policy applies. We therefore evaluate the risk by its threshold-free
ranking of disruptive and non-disruptive discharges using only the
information available before a required deadline (Deadline AUPRC). We
use the area under the precision--recall curve because disruptive
discharges are a minority, so that the area under the receiver
operating characteristic curve can be uninformative
\citep{saito2015precision}, and we keep detection and warning time
separate rather than combining them into a single quantity such as the
area under the warning-time characteristic curve
\citep{keith2024riskaware}. In operation, however, the alarm policy
must fix its threshold in advance, and what matters is whether
disruptions are detected in time while false alarms during normal
operation remain rare. We therefore also evaluate alarm policies whose
thresholds are calibrated on validation data, by the Deadline F1 and
the false-alarm rate per 24 hours of normal operation on unseen test
discharges. Separating the two evaluations reveals whether a difference
in alarm performance stems from the risk itself or from the
calibration of its threshold.

The main contributions of this work are as follows:

\begin{enumerate}
\item We introduce two horizon-focused prediction targets,
horizon-truncated survival analysis and fixed-horizon EEP with TLS, into
tokamak disruption prediction, and we compare them systematically with
BC and SA under a common protocol on three tokamaks.
\item We isolate the effect of temporal label smoothing by comparing
TLS with hard fixed-horizon targets placed at either end of its
smoothing range, with the training data, encoder, and optimization
fixed.
\item We propose an evaluation protocol that separates the risk, through
its threshold-free ranking at an operational deadline, from the alarm
policy, through the performance of validation-calibrated thresholds.
\end{enumerate}

%% file: manuscript/related_work.tex
\section{Related Work}\label{sec:related_work}

\subsection{Machine Learning for Disruption Detection}
\label{sec:related-disruption}

Machine-learning-based disruption prediction has evolved from device-specific classifiers built on manually selected plasma quantities to deep sequence models and, more recently, standardized multi-machine evaluation. Early studies demonstrated that disruption precursors could be inferred from diagnostic signals using cross-tokamak neural networks, support vector machines, fuzzy-logic and regression-tree systems, and unsupervised or manifold-based representations \citep{windsor2005cross,cannas2007svm,murari2008prototype,murari2009unbiased,ratta2010advanced}. Despite their architectural diversity, most of these methods formulated disruption prediction as fixed-window binary classification.

Most of these predictors label each time slice of a disruptive shot
as positive when it lies within a prescribed class time
$\Delta_{\mathrm{class}}$ of the disruption, and raise an alarm when
the predicted probability satisfies a threshold or hysteresis rule.
This formulation imposes an abrupt, user-defined boundary between
non-disruptive and pre-disruptive states. However, the relevant
precursor timescale depends on the underlying disruption mechanism. For
example, locked-mode precursors may emerge over tens of milliseconds,
whereas impurity accumulation can evolve over hundreds of milliseconds.
Consequently, a single $\Delta_{\mathrm{class}}$ cannot capture the
temporal structure of all disruption processes, nor does it necessarily
coincide with their physical onset \citep{keith2024riskaware}. We
relate this formulation to the prediction targets compared in this
study in \cref{sec:methods-targets}.

Random-forest predictors trained on large DIII-D and Alcator C-Mod databases were subsequently developed and deployed in real-time operation \citep{rea2018investigations,rea2019realtime}. Later research largely replaced manually engineered features with recurrent neural networks (RNNs), temporal convolutional networks (TCNs), spatiotemporal convolutional models, and Transformers that learn temporal representations directly from diagnostic signals \citep{katesharbeck2019predicting,churchill2020deep,zhu2021hybrid,aymerich2022profiles}. DisruptionBench compared a random forest, the hybrid deep learner, a Transformer, and a continuous convolutional neural network (CCNN) \citep{arnold2023continuous} across Alcator C-Mod, DIII-D, and EAST under common data splits and alarm logic \citep{spangher2025disruptionbench}. Long-context studies have shown that predictive information is not limited to the immediate pre-disruption interval but is also present much earlier in the plasma evolution, motivating the use of sequence encoders capable of preserving long-range dependencies \citep{spangher2025disruptionbench}.

More recently, disruption prediction has increasingly been formulated within the framework of survival analysis \citep{olofsson2018hazard,tinguely2019survival,keith2024riskaware}. A key advantage of this framework is its principled treatment of censored observations and its ability to explicitly model the time remaining until disruption. \citet{olofsson2018hazard} modeled the hazard function for tearing-mode onset, while \citet{tinguely2019survival} combined random-forest predictions with conditional Kaplan--Meier estimation to obtain horizon-dependent disruption risks. More recently, \citet{keith2024riskaware} compared conditional Kaplan--Meier estimation with Cox proportional hazards, Deep Cox proportional hazards, and DSM.

These survival-based studies model the time to disruption over its full range. In parallel, clinical early-warning research has developed survival and fixed-horizon formulations that focus on the time window relevant to an alarm, as reviewed below; their potential for disruption prediction has not yet been assessed.

\subsection{EEP}
\label{sec:related-eep}

Survival analysis models the time to an event while accounting for
right censoring. Classical approaches include the Kaplan--Meier
estimator and Cox proportional hazards, and later work introduced
nonlinear, discrete-time, and fully parametric deep survival models
\citep{kaplan1958nonparametric,cox1972regression,faraggi1995neural,
gensheimer2019scalable,lee2018deephit,kvamme2019time,nagpal2021dsm,
chen2024introduction}. Dynamic prediction updates these estimates as
longitudinal measurements accumulate, either by landmarking
\citep{vanhouwelingen2007landmarking} or with recurrent deep models
\citep{lee2020dynamicdeephit}.

EEP developed largely in clinical early-warning systems, where
multivariate time series from electronic health records are used to
predict adverse events, such as acute kidney injury, circulatory
failure, or decompensation, before they occur
\citep{tomasev2019aki,hyland2020circulatory,harutyunyan2019multitask,yeche2021hirid}.
These systems typically predict, at every time step, whether the event
will occur within a fixed horizon, so that the prediction is aligned
with the decision to raise an alarm, and they evaluate the resulting
alarms at the event level rather than per time step.

Subsequent work examined the training target itself.
\citet{yeche2023tls} related fixed-horizon EEP to dynamic survival
analysis and proposed TLS, which replaces the abrupt fixed-horizon
label with a target that varies smoothly with the time to the event.
\citet{yeche2024dsa} adapted dynamic survival analysis to EEP by
restricting the survival objective to the horizon of interest and
localizing risk within it, and they used this within-horizon
information to prioritize alarms.

The survival-based disruption predictors discussed above, in contrast,
model the residual time over its full range. The targets of binary
classification, full-distribution and horizon-truncated survival
analysis, and fixed-horizon EEP compared in this study are defined in
\cref{sec:methods}.

%% file: manuscript/method.tex
\section{Methods}
\label{sec:methods}
\label{sec:methods-overview}

We first fix the notation (\cref{sec:methods-setup}) and contrast the
three prediction targets used for disruption prediction
(\cref{sec:methods-targets}). We then follow the progression from
modeling the full time to disruption to modeling only the risk within
the horizon: survival analysis (\cref{sec:methods-sa}), the
horizon-truncated survival objective (\cref{sec:methods-lsurv}), and
fixed-horizon EEP with TLS (\cref{sec:methods-tls}). Finally, we describe
the causal encoder shared by the neural predictors
(\cref{sec:methods-encoder}).

\subsection{Problem Setup}
\label{sec:methods-setup}

Consider shot $i\in\{1,\ldots,N\}$, represented by a multivariate
diagnostic sequence
$\mathcal{X}_{i}=(\mathbf{x}_{i,0},\mathbf{x}_{i,\Delta_x},\ldots,\mathbf{x}_{i,L_i\Delta_x})$
with $\mathbf{x}_{i,t}\in\mathbb{R}^{d}$, where $L_i$ is the index of the final valid diagnostic sample of shot
$i$, so that the sequence ends at physical time $L_i\Delta_x$. The
sequence is sampled on the diagnostic input grid
$\Delta_x=5\,\mathrm{ms}$. We use $t$ for physical landmark time, so
$t\in\{0,\Delta_x,\ldots,L_i\Delta_x\}$. Let $T_i$ be the physical
disruption time and $C_i$ the physical censoring time, both measured in
milliseconds. The observed endpoint and event indicator are
$Y_i=\min(T_i,C_i)$ and $\delta_i=\mathbb{I}\{T_i\leq C_i\}$. All temporal quantities in the objectives below, including $t$, $Y_i$,
$R_{i,t}$, and the training horizon $H$, are physical times rather than
numbers of diagnostic input steps.

At each causal prediction origin, or landmark, $t<Y_i$, the observed
residual follow-up time is $R_{i,t}=Y_i-t$. If $\delta_i=1$, $R_{i,t}$ is the observed time remaining until
disruption. If $\delta_i=0$, the discharge ends without a disruption
after $R_{i,t}$. Survival analysis treats this end as right censoring,
which implies only that the disruption time would exceed $R_{i,t}$ if
the discharge continued. Because a discharge that has ended cannot
disrupt afterwards, the end of a non-disruptive shot can instead be
treated as a terminal event that precludes disruption. The SA models of
\cref{sec:methods-sa} use the first view, and the EEP targets and
survTLS (\cref{sec:methods-lsurv,sec:methods-tls}) use the second.

At landmark $t$, the measurements available to a causal predictor are
the full observed prefix
$\mathcal{X}_{i,0:t}=(\mathbf{x}_{i,0},\ldots,\mathbf{x}_{i,t})$, which excludes all future measurements
$\mathbf{x}_{i,t+\Delta_x:L_i\Delta_x}$. We denote by $\mathcal{I}_i$ the
set of valid landmarks of shot $i$ used for training
(\cref{app:shot-partitions}).

\subsection{Three Prediction Targets}
\label{sec:methods-targets}

\begin{figure}[t]
    \centering
    \includegraphics[width=\columnwidth]{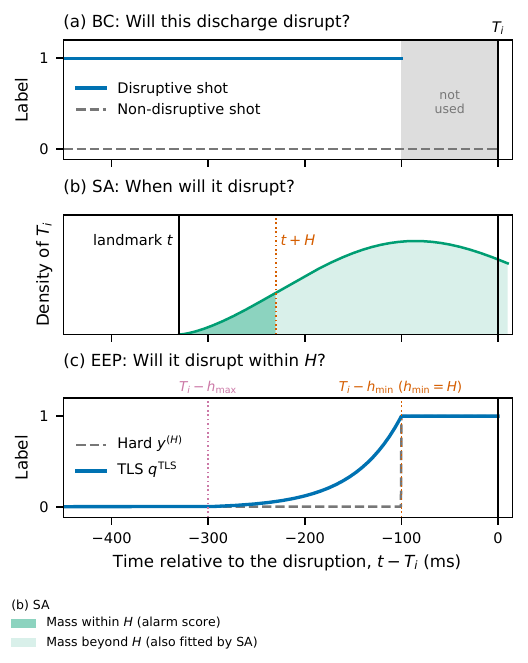}
    \caption{
        Training targets for a disruptive shot. (a) BC assigns one label
        to the whole discharge. (b) SA models the distribution of the
        disruption time at every landmark, including the mass beyond
        $t+H$. (c) EEP labels every landmark by whether the disruption
        occurs within the horizon $H$; TLS replaces the hard boundary at
        $T_i-H$ by a target that decreases from one at $T_i-h_{\min}$,
        with $h_{\min}=H$, to zero at $T_i-h_{\max}$. The values of $H$
        and of the smoothing are illustrative.
    }
    \label{fig:prediction-targets}
\end{figure}

Disruption predictors differ in the risk they are trained to output.
\Cref{fig:prediction-targets} contrasts the training targets of binary
classification (BC), survival analysis (SA), and fixed-horizon EEP for a
disruptive shot. The three targets answer different questions, and
they differ in how much they say about when the disruption occurs.

\paragraph{Binary classification: is this discharge disruptive?}
BC labels whether a discharge, or a state within it, is disruptive. In
its simplest form, every landmark of shot $i$ receives the shot outcome
$d_{i,t}=\delta_i$,
and the model estimates
$\Pr_{\theta}(d_{i,t}=1\mid\mathcal{X}_{i,0:t})$. Alternatively, a
pre-disruptive state can be annotated from the physical onset of a
precursor. BC is a simple and widely used problem, but its label does
not specify when the disruption occurs relative to the landmark. A
shot-level label also asks the model to flag early parts of a
disruptive discharge that may not yet contain any precursor, and
an onset-based annotation depends on the disruption mechanism
\citep{keith2024riskaware}.

\paragraph{Survival analysis: when will the discharge disrupt?}
SA models the conditional distribution of the disruption time while
accounting for right censoring \citep{kalbfleisch2002statistical}. It
provides the time to disruption directly and the risk at any horizon
from a single model \citep{tinguely2019survival,keith2024riskaware}.

\paragraph{Fixed-horizon EEP: will the discharge disrupt within $H$?}
EEP labels every landmark from its residual time and an operational
prediction horizon $H$: the label is positive when the disruption
occurs within the next $H$ (\cref{eq:method-eep-target-mask}), and the
model estimates the probability of this event. No annotation of a
pre-disruptive state is required, and learning concentrates on the
horizon that matters for the control action. Conventional class-time classifiers \citep{keith2024riskaware} label the
last $\Delta_{\mathrm{class}}$ of a disruptive shot as positive; when
$\Delta_{\mathrm{class}}$ is chosen as an operational prediction
horizon, this target coincides with the hard EEP target, and we
classify such models as EEP.

\paragraph{Two axes.}
The targets thus differ along two axes: whether they model the
residual time (survival modeling), and whether learning is restricted
to the operational horizon (horizon focus). SA models the residual time
over its full range without horizon focus. The horizon-truncated
survival objective of \cref{sec:methods-lsurv} keeps survival modeling
but restricts it to the horizon, and EEP focuses on the horizon without
modeling the residual time.

\subsection{Survival Analysis}
\label{sec:methods-sa}

At landmark $t$, an SA model with parameters $\Theta$ defines a
conditional density $f_{\Theta}(\tau\mid\mathbf{z}_{i,t})$ of the
residual time and the survival function
$S_{\Theta}(\tau\mid\mathbf{z}_{i,t})=\int_{\tau}^{\infty}f_{\Theta}(u\mid\mathbf{z}_{i,t})\,\mathrm{d}u$,
where $\mathbf{z}_{i,t}$ is the encoded prefix
(\cref{sec:methods-encoder}). Under conditional independent censoring,
the censored likelihood over causal landmarks gives the objective
\citep{chen2024introduction}
\begin{multline}
    \mathcal{L}_{\mathrm{SA}}(\Theta)
    =
    -\sum_i\sum_{t<Y_i}
    \Bigl[
        \delta_i\log f_{\Theta}\left(R_{i,t}\mid\mathbf{z}_{i,t}\right)
    \\
        +(1-\delta_i)\log S_{\Theta}\left(R_{i,t}\mid\mathbf{z}_{i,t}\right)
    \Bigr].
    \label{eq:method-dsm-loss}
\end{multline}
The alarm score for horizon $H$ is the predicted probability of a
disruption within $H$, $1-S_{\Theta}(H\mid\mathbf{z}_{i,t})$. We use
two SA models evaluated in previous disruption studies
\citep{keith2024riskaware}. DSM \citep{nagpal2021dsm} parameterizes
$f_{\Theta}$ as a mixture of $M$ Weibull distributions whose weights and
parameters are produced from $\mathbf{z}_{i,t}$, and optimizes a lower
bound of \cref{eq:method-dsm-loss}. DCPH \citep{faraggi1995neural} is a
deep Cox proportional-hazards model whose hazard is a baseline hazard
multiplied by $\exp(g_{\Theta}(\mathbf{z}_{i,t}))$. It is trained with
the Cox partial likelihood, and its survival function uses a baseline
cumulative hazard $\Lambda_0$ estimated on the training data. Its horizon
risk $1-\exp[-\Lambda_0(H)e^{g_{\Theta}(\mathbf{z}_{i,t})}]$ is a strictly
increasing function of $g_{\Theta}(\mathbf{z}_{i,t})$ whenever
$\Lambda_0(H)>0$, so the ranking of landmarks does not depend on $H$.

\subsection{Horizon-Truncated Survival Objective (survTLS)}
\label{sec:methods-lsurv}

Learning the complete event-time distribution with
$\mathcal{L}_{\mathrm{SA}}$ requires fitting residual event times over
the full observed range, including events far beyond the horizon
relevant to the alarm. On long and finely sampled sequences, these
distant and weakly predictable events can make optimization
unnecessarily difficult. A first way to focus learning on the
operational horizon keeps the survival model but truncates its
objective at a physical training horizon $H$. survTLS
\citep{yeche2024dsa} estimates a sequence of discrete hazards over
future-time bins and fits them only within the horizon, so that it
still localizes the risk in time within $H$.

Let $\delta_h$ be the output-bin width and $T_{\max}\geq H$ the time
range of the outputs, so that the head has $K=T_{\max}/\delta_h$
outputs. The number of bins within the training horizon is
$K_H=\lceil H/\delta_h\rceil$. Bin $k\in\{1,\ldots,K\}$ is centered at $k\delta_h$ and covers
$B_k=\bigl((k-\tfrac{1}{2})\delta_h,(k+\tfrac{1}{2})\delta_h\bigr]$. The
hazard head predicts
$\lambda_{\theta}(k\mid\mathbf{z}_{i,t})=\sigma(b_k(\mathbf{z}_{i,t}))$,
the conditional probability that the event occurs in $B_k$ given that it
has not occurred in an earlier bin, where $b=(b_1,\ldots,b_K)$ is a
neural prediction head.

The one-hot residual-time target is replaced by a smoothed event-time
probability mass $\widetilde{f}_{i,t}(k)$, obtained by integrating over
each bin a Gaussian centered at $R_{i,t}$ with standard deviation
$R_{i,t}/\ell$, where $\ell$ is a dimensionless length scale; distant
events thus receive less certain temporal localization. The mass below
the first bin, including negative times, is assigned to $k=1$, and the
mass above $T_{\max}$ is assigned to $k=K$, so that
$\sum_{k=1}^{K}\widetilde{f}_{i,t}(k)=1$; the precise expressions are
given in \cref{app:survtls-implementation}. Normal termination is
treated as a terminal event (\cref{sec:methods-setup}): for landmarks of
non-disruptive shots, $\widetilde{f}_{i,t}(k)=0$ for every $k$, so that
every bin within the horizon is negative even if the discharge ends
earlier. Define the soft at-risk weight and the target hazard by
\begin{gather}
    u_{i,t,k}
    =
    1-\sum_{j=1}^{k-1}\widetilde{f}_{i,t}(j),
    \qquad
    \widetilde{\lambda}_{i,t}(k)
    =
    \frac{\widetilde{f}_{i,t}(k)}{u_{i,t,k}},
    \label{eq:method-survtls-target-hazard}
\end{gather}
so that non-disruptive landmarks receive zero target hazards with unit
weight. Because $u_{i,t,k}\widetilde{\lambda}_{i,t}(k)=\widetilde{f}_{i,t}(k)$
and $u_{i,t,k}(1-\widetilde{\lambda}_{i,t}(k))=1-\sum_{j\leq k}\widetilde{f}_{i,t}(j)$,
the weighted terms of \cref{eq:method-lsurv} summed over all $K$ bins
give the expected discrete-time log-likelihood under the smoothed
event-time distribution,
\begin{multline}
    \sum_{k=1}^{K}
    \Bigl[
        \widetilde{f}_{i,t}(k)\log\lambda_{\theta,k}
    \\
        +\Bigl(1-\sum_{j\leq k}\widetilde{f}_{i,t}(j)\Bigr)
        \log\left(1-\lambda_{\theta,k}\right)
    \Bigr],
    \label{eq:method-survtls-likelihood}
\end{multline}
where $\lambda_{\theta,k}=\lambda_{\theta}(k\mid\mathbf{z}_{i,t})$. A sharp
event therefore keeps its full weight in its own bin. The horizon-truncated objective uses only the first $K_H$ bins
and is averaged over the valid landmarks of each shot and then over the
$N$ training shots:
\begin{multline}
    \mathcal{L}_{\mathrm{surv}}^{(H)}
    =
    -\frac{1}{N}\sum_{i=1}^{N}\frac{1}{|\mathcal{I}_i|}
    \sum_{t\in\mathcal{I}_i}\sum_{k=1}^{K_H}
    u_{i,t,k}
    \\
    \times\Bigl[
        \widetilde{\lambda}_{i,t}(k)
        \log\lambda_{\theta}\left(k\mid\mathbf{z}_{i,t}\right)
    \\
        +\left(1-\widetilde{\lambda}_{i,t}(k)\right)
        \log\left(1-\lambda_{\theta}\left(k\mid\mathbf{z}_{i,t}\right)\right)
    \Bigr].
    \label{eq:method-lsurv}
\end{multline}
The targets are computed over all $K$ bins and are not renormalized
after the truncation to $K_H$ bins.

Restricting the objective to $k\leq K_H$ avoids fitting event times
beyond the training horizon, and the ordered hazard profile provides a
fine-grained representation of event imminence within $H$. At inference
time, the hazards are aggregated into an alarm score that gives greater
weight to shorter future offsets (\cref{app:alarm-scores}).

\subsection{Fixed-Horizon EEP with TLS}
\label{sec:methods-tls}

survTLS still models when the disruption occurs within the horizon.
Fixed-horizon EEP goes one step further: it gives up modeling the
residual time and directly learns the probability of a disruption
within the next $H$, which aligns the objective with the early-warning
task \citep{yeche2023tls}.

The hard fixed-horizon target is
\begin{equation}
    y_{i,t}^{(H)}
    =
    \mathbb{I}
    \left\{
        \delta_i=1
        \;\wedge\;
        0<R_{i,t}\leq H
    \right\}.
    \label{eq:method-eep-target-mask}
\end{equation}
Normal termination is treated as a terminal event: every landmark of a
non-disruptive shot is negative for every $H$, including landmarks
within $H$ of the end of the discharge, because no disruption can occur
after the end. The label is therefore observed at every valid landmark
$t\in\mathcal{I}_i$, and the losses below use all valid landmarks
without a censoring mask. A scalar prediction head outputs
$p_{\theta}^{(H)}(\mathbf{z}_{i,t})\in(0,1)$. When it is trained on the
hard target, it estimates the probability that a disruption occurs
within the next $H$ milliseconds and before the discharge ends. This
quantity differs from the horizon risk of the SA models, which treat
the end of a non-disruptive shot as censoring and thus describe the
disruption time of a discharge that would continue.

The hard target has two weaknesses. It switches abruptly at
$T_i-H$, although the predictive signal does not change abruptly at
this boundary, and only the last $H$ of each
disruptive shot is positive, which causes severe label imbalance. TLS
\citep{yeche2023tls} replaces the step with a target that decreases
smoothly with the residual time. It reduces confidence where the signal
is weak or ambiguous and concentrates learning on the comparatively
rare event-proximal landmarks.

Let $h_{\min}<h_{\max}$ delimit the smoothing range in milliseconds,
$\gamma>0$, in $\mathrm{ms}^{-1}$, control its curvature, and
$D_{\mathrm{TLS}}=h_{\max}-h_{\min}$. Because $\gamma$ multiplies a
time, its value depends on the time unit; we use
$\gamma=0.025\,\mathrm{ms}^{-1}$ ($25\,\mathrm{s}^{-1}$). The exponential TLS smoothing
profile is
\begin{equation}
    \rho_{\gamma}(\tau)
    =
    \begin{cases}
        1,
        & \tau\leq h_{\min},
        \\[1mm]
        0,
        & \tau\geq h_{\max},
        \\[1mm]
        \dfrac{
            e^{-\gamma(\tau-h_{\min})}
            -
            e^{-\gamma D_{\mathrm{TLS}}}
        }{
            1-e^{-\gamma D_{\mathrm{TLS}}}
        },
        & \text{otherwise}.
    \end{cases}
    \label{eq:method-tls-smoothing}
\end{equation}
The smoothed target at landmark $t$ is
$q_{i,t}^{\mathrm{TLS}}=\delta_i\rho_{\gamma}(R_{i,t})$.
Because $\rho_{\gamma}$ decreases from one to zero between $h_{\min}$ and
$h_{\max}$, the TLS target is bounded by the hard targets of
\cref{eq:method-eep-target-mask} at these two horizons:
\begin{equation}
    y_{i,t}^{(h_{\min})}
    \leq
    q_{i,t}^{\mathrm{TLS}}
    \leq
    y_{i,t}^{(h_{\max})}.
    \label{eq:method-tls-bounds}
\end{equation}
The three targets coincide outside $(h_{\min},h_{\max})$, so Hard EEP at
$H=h_{\min}$ or $H=h_{\max}$ differs from TLS only in the values assigned
within the smoothing range.

The TLS objective is the binary cross-entropy averaged over the valid
landmarks of each shot and then over the $N$ training shots:
\begin{multline}
    \mathcal{L}_{\mathrm{TLS}}^{(H)}
    =
    -\frac{1}{N}\sum_{i=1}^{N}\frac{1}{|\mathcal{I}_i|}
    \sum_{t\in\mathcal{I}_i}
    \Bigl[
        q_{i,t}^{\mathrm{TLS}}\log p_{\theta}^{(H)}(\mathbf{z}_{i,t})
    \\
        +\left(1-q_{i,t}^{\mathrm{TLS}}\right)
        \log\left(1-p_{\theta}^{(H)}(\mathbf{z}_{i,t})\right)
    \Bigr].
    \label{eq:method-tls-loss}
\end{multline}

Replacing $q_{i,t}^{\mathrm{TLS}}$ in \cref{eq:method-tls-loss} by the
hard target $y_{i,t}^{(H)}$ gives the hard-label EEP objective, which we
refer to as Hard EEP. When trained on the TLS target instead, the
output estimates the conditional expectation of
$q_{i,t}^{\mathrm{TLS}}$ given the prefix. Because
$q_{i,t}^{\mathrm{TLS}}$ takes intermediate values for residual times
between $h_{\min}$ and $h_{\max}$, this quantity is not the probability
of a disruption within a fixed horizon. By \cref{eq:method-tls-bounds},
however, it lies between the probabilities of a disruption within
$h_{\min}$ and within $h_{\max}$, so its meaning as a risk is fixed by
the two bounds. We therefore treat the TLS output as a smoothed,
horizon-aware risk score and use it to rank landmarks and to trigger
alarms. At inference
time, the model monitors $p_{\theta}^{(H)}(\mathbf{z}_{i,t})$, and an
alarm is raised when this score exceeds a threshold calibrated on
validation data.

\subsection{Shared Causal Encoder}
\label{sec:methods-encoder}

All neural predictors except CCNN are conditioned on a common causal
feature representation. The encoder separates a recurrent history branch from
a recent-window branch that directly retains high-resolution
measurements.

At landmark $t$, the history branch has processed
$\mathcal{X}_{i,0:t}$ causally and maintains the recurrent
representation
$\mathbf{z}_{i,t}^{\mathrm{hist}}=g_{\mathrm{hist}}(\mathcal{X}_{i,0:t})$.
The recurrent state is propagated through the complete causal prefix,
while gradients are truncated to the most recent 256 input steps during
backpropagation through time (BPTT). The forward state can therefore
carry information from before the truncation boundary, but
gradient-based credit assignment does not extend through the complete
prefix.

The recent-window branch directly retains the ordered measurements in
the most recent $L_{\mathrm{win}}$ diagnostic input samples. Defining
$\mathbf{x}_{i,t'}=\mathbf{0}$ for $t'<0$, the causal recent window and
its representation are
$\mathbf{w}_{i,t}=[\mathbf{x}_{i,t-(L_{\mathrm{win}}-1)\Delta_x};\ldots;\mathbf{x}_{i,t}]\in\mathbb{R}^{L_{\mathrm{win}}d}$
and $\mathbf{z}_{i,t}^{\mathrm{win}}=g_{\mathrm{win}}(\mathbf{w}_{i,t})$.
Unlike the history representation, the recent-window representation
bypasses recurrent compression of the recent measurements and
therefore preserves short-timescale precursor dynamics explicitly.

The two representations are concatenated as
$\mathbf{z}_{i,t}=[\mathbf{z}_{i,t}^{\mathrm{hist}};\mathbf{z}_{i,t}^{\mathrm{win}}]$.
A single-branch model uses only the corresponding component of
$\mathbf{z}_{i,t}$. In the reported experiments,
$g_{\mathrm{hist}}$ is instantiated as a GRU and
$g_{\mathrm{win}}$ as an MLP applied to the flattened recent window.
These are implementation choices for the two information-preservation
mechanisms; the prediction objectives below are independent of the
selected encoder architecture.

%% file: manuscript/experiments.tex


\section{Experiments}
\label{sec:experiments}

\subsection{Experimental Design}
\label{sec:experimental-design}

We conduct three studies on DIII-D, Alcator C-Mod, and EAST, treating
each tokamak as a separate within-device task.
\begin{enumerate}
\item \emph{Predictor comparison.} We compare predictors that differ in
the quantity they are trained to predict: binary classification (BC),
full-distribution survival analysis (SA), and fixed-horizon EEP
(\cref{sec:experiment-models}). All neural sequence
models except CCNN share the causal encoder of
\cref{sec:methods-encoder}; CCNN and RF use their own feature
extraction.
\item \emph{Training horizon and evaluation deadline.} Each method
selects its training horizon $H$, or for TLS its smoothing bounds
$(h_{\min},h_{\max})$, on validation data
(\cref{sec:experiment-optimization}). Hard EEP is trained at
$H=h_{\min}$ of the selected TLS bounds, so that it differs from TLS
only in the smoothing, and Hard EEP at $H=h_{\max}$ is reported as a
second reference (\cref{sec:results-smoothing}). Every selected model is
evaluated at the deadlines $w\in\{10,40,100,300\}\,\mathrm{ms}$. For
TLS, we additionally fix each upper bound
$h_{\max}\in\{20,40,80,120,240,360\}\,\mathrm{ms}$, select its lower
bound on the monitor split, and train it with three seeds, which gives
an $h_{\max}\times w$ grid.
\item \emph{Encoder ablation.} For TLS, we fix the selected bounds and
training cohort and remove either the history branch or the
recent-window branch of the encoder. The conclusions of this ablation
therefore concern the encoder used with the TLS objective.
\end{enumerate}
Within each study, all settings other than the studied factor are
fixed.

\subsection{Datasets and Preprocessing}
\label{sec:experiment-datasets}

We use the public DIII-D, Alcator C-Mod, and EAST databases from
DisruptionBench \citep{spangher2025disruptionbench}. Each shot comprises
a multivariate diagnostic sequence and a disruptive or
non-disruptive endpoint. For DSM and DCPH, non-disruptive shots are
right-censored at their last valid observation; for the EEP models and
survTLS, their landmarks are negative (\cref{sec:methods-tls}).

Nine plasma quantities are common to the three databases. They describe
magnetohydrodynamic and equilibrium state, density-limit proximity,
plasma shaping, current tracking, and loop-voltage evolution; symbols
and units are listed in \cref{app:diagnostic-quantities}. The DIII-D
and EAST databases each provide three additional device-specific
signals, which the neural sequence models use as additional input
channels. The C-Mod input contains three machine-indicator channels instead,
which are constant within the within-device task and therefore carry no
information.

\begin{table}[t]
    \centering
    \small
    \setlength{\tabcolsep}{4.0pt}
    \caption{
        DisruptionBench database composition before resampling;
        parentheses give disruptive-shot fractions.
    }
    \label{tab:dataset-composition}
    \begin{tabular}{lrrr}
        \toprule
        Device & Shots & Disruptions & Native interval \\
        \midrule
        C-Mod  & 4,510  & 975 (21.6\%)   & $5\,\mathrm{ms}$   \\
        \midrule
        DIII-D & 8,608  & 1,244 (14.5\%) & $25\,\mathrm{ms}$  \\
        \midrule
        EAST   & 14,347 & 4,508 (31.4\%) & $100\,\mathrm{ms}$ \\
        \bottomrule
    \end{tabular}
\end{table}

Signals are placed on a causal $5\,\mathrm{ms}$ grid by forward-filling
only previously observed values; this common diagnostic input-grid
interval is denoted by $\Delta_x=5\,\mathrm{ms}$. We use shots whose
prepared sequences contain between 15 and 2,048 landmarks
($75\,\mathrm{ms}$ to about $10\,\mathrm{s}$), and we disable
absolute-time and short-time Fourier transform (STFT) features. Each
channel is standardized with statistics estimated on the training
partition. Shot-level training, calibration, monitoring, and test
fractions are $60\%$, $10\%$, $10\%$, and $20\%$, assigned once with a
fixed seed and stratified by shot outcome; partitioning precedes prefix
generation. The disruption time $T_i$ is taken from the
DisruptionBench databases and is resolved on the $5\,\mathrm{ms}$ grid.
The resulting partition sizes, the observation time of non-disruptive
shots, the eligibility rules, and the temporal resolution of $T_i$ are
given in \cref{app:shot-partitions}.

\subsection{Compared Models}
\label{sec:experiment-models}

\Cref{tab:compared-models} lists the compared predictors, grouped by the
prediction targets of \cref{sec:methods-targets}. All neural predictors
except CCNN use the shared encoder of \cref{sec:methods-encoder};
implementation details and method-specific hyperparameters are listed in
\cref{app:optimization,app:method-settings}.

\begin{table}[t]
    \centering
    \scriptsize
    \setlength{\tabcolsep}{1.6pt}
    \caption{
        Compared predictors, grouped by prediction target
        (\cref{sec:methods-targets}). ``Shared'' denotes the causal
        encoder of \cref{sec:methods-encoder}.
    }
    \label{tab:compared-models}
    \begin{tabular}{llll}
        \toprule
        Method & Target & Label or objective & Model \\
        \midrule
        CCNN & BC & Shot label $d_{i,t}$ & Conv.\ net \\
        DSM & SA & $\mathcal{L}_{\mathrm{SA}}$ (Weibull mix.) & Shared \\
        DCPH & SA & Cox partial lik. & Shared \\
        survTLS & SA ($\leq H$) & $\mathcal{L}_{\mathrm{surv}}^{(H)}$ & Shared \\
        RF & EEP & Hard $y^{(H)}_{i,t}$ & Forest \\
        Hard EEP & EEP & Hard $y^{(h_{\min})}_{i,t}$ & Shared \\
        TLS & EEP & Smoothed $q^{\mathrm{TLS}}_{i,t}$ & Shared \\
        \bottomrule
    \end{tabular}
\end{table}

\paragraph{BC.}
CCNN \citep{arnold2023continuous}, the continuous convolutional network
evaluated in DisruptionBench \citep{spangher2025disruptionbench}, is
trained as a sequence-to-label classifier on the shot label
$d_{i,t}=\delta_i$, with each disruptive training sequence truncated at
$T_i-H$; its score at landmark $t$ is its prediction for the prefix
ending at $t$. Because its training settings and evaluation protocol
differ from the original ones (\cref{app:ccnn-differences}), the
results are not directly comparable with the published intra-machine
C-Mod AUC of approximately 0.97.

\paragraph{SA.}
DSM uses Auton-Survival,%
\footnote{Auton-Survival:
\url{https://github.com/autonlab/auton-survival}}
the implementation used in previous risk-aware disruption-prediction
research \citep{keith2024riskaware}, and is trained on every valid
landmark (\cref{app:method-settings}). For DCPH, each risk set of the
partial likelihood is formed from all valid landmarks in a mini-batch
of shots. survTLS follows the horizon-truncated objective and
risk-localization construction of \citet{yeche2024dsa}.

\paragraph{EEP.}
TLS follows the authors' official target construction with the
exponential smoothing profile of \cref{eq:method-tls-smoothing}.%
\footnote{TLS: \url{https://github.com/ratschlab/tls}}
Hard EEP is trained on the hard target with $H=h_{\min}$ of the selected
TLS bounds and shares the training cohort, encoder, and optimization of
TLS, so that the two differ only in the smoothing of the target. RF is
a random-forest classifier, the model family used in earlier real-time
disruption predictors \citep{rea2018investigations,rea2019realtime},
trained on the hard target $y^{(H)}_{i,t}$ with 63 causal window
features of the nine common signals (\cref{app:method-settings}); it
therefore differs from Hard EEP in both its input and its model class.

\subsection{Optimization and Model Selection}
\label{sec:experiment-optimization}

All neural models share the optimizer, schedule, and stopping rule
(\cref{app:optimization,app:method-settings}), and for each candidate
configuration we retain the checkpoint with the largest monitor-split
Deadline AUROC at $w=40\,\mathrm{ms}$. CCNN, DSM, and DCPH search
$H\in\{20,40,80,120\}\,\mathrm{ms}$, and RF and survTLS search
$H\in\{20,40,80,120,240,360\}\,\mathrm{ms}$. DSM additionally searches
the number of mixture components $M\in\{1,3\}$, and survTLS the
output-bin width and learning rate. TLS searches the 22 smoothing
intervals $(h_{\min},h_{\max})$ of \cref{tab:tls-candidates}. For DCPH,
$H$ changes only which short disruptive shots are eligible for
training, not the ranking produced by a given model
(\cref{app:dcph-horizon}). After threshold calibration
(\cref{sec:alarm-calibration}), the configuration with the largest
monitor-split Deadline F1 of seed 0 is selected and frozen for seeds 1
and 2 (\cref{app:evaluation-calibration-selection}). Test data are not
used for any selection.

\subsection{Evaluation Protocol}
\label{sec:evaluation-protocol}
\label{sec:evaluation-metric-overview}

Following \cref{sec:introduction}, we evaluate separately the risk
output by each predictor and the alarm policy built on it. A useful
alarm must be raised sufficiently far in advance of a disruption to
allow control or mitigation measures to be implemented. We denote this
required lead time by $w$ and use
$w=40\,\mathrm{ms}$ as the primary deadline, following the
DisruptionBench setting for the ITER disruption-mitigation response
\citep{spangher2025disruptionbench}. Because the appropriate deadline
depends on the available control action \citep{keith2024riskaware}, the
horizon study additionally evaluates $w\in\{10,100,300\}\,\mathrm{ms}$.
$H$ is the physical training horizon used to construct the training
target, whereas $w$ is the required alarm lead time used only to define
the evaluation deadline; the two quantities need not be equal.

\paragraph{Ranking metric.}
\label{sec:evaluation-warning-auroc}
Deadline AUPRC evaluates the risk: it measures whether disruptive shots
receive higher scores than non-disruptive shots, using only information
available by the deadline. \citet{keith2024riskaware} used AUROC for
this purpose; because disruptive shots are a minority, we use AUPRC, as
in EEP evaluation \citep{saito2015precision,yeche2023tls}, and report
Deadline AUROC in \cref{app:supplementary-results}.

\paragraph{Alarm metrics.}
\label{sec:evaluation-deadline}
\label{sec:evaluation-event-alarm}
\label{sec:evaluation-far}
The alarm policy is evaluated at the event level, following EEP
evaluation
\citep{yeche2023tls,yeche2024dsa}, Deadline Recall is the proportion of
disruptions for which at least one alarm is raised no later than $w$
before the disruption, and Alarm Precision is the proportion of emitted
alarms that are useful. At most one true alarm is credited per
disruptive shot; all normal-shot, late, and duplicate alarm episodes
are counted as false alarms. Deadline F1 is the harmonic mean of the
two, and FAR/24h is the number of alarm events in non-disruptive shots
normalized to 24 hours of normal operation. Definitions are given in
\cref{app:evaluation-definitions}.

\paragraph{Alarm policy.}
\label{sec:alarm-construction}
\label{sec:alarm-calibration}
Each method's score sequence (\cref{app:alarm-scores}) is post-processed
with the fixed settings of \cref{app:checkpoint-selection} and
thresholded. The threshold is selected on the calibration split so that
the normal-shot false-positive rate (FPR), the proportion of
non-disruptive shots with at
least one alarm, is approximately $1\%$. Together with the checkpoint
and configuration selected on the monitor split, this defines a
validation-selected fixed policy that is applied unchanged to the test
split. To separate calibration from ranking, we also sweep the
threshold on the test data and trace Deadline Recall against the
realized FAR/24h. These Recall--FAR curves use test labels and are
descriptive only.
\phantomsection\label{sec:evaluation-matched-far}

\paragraph{Uncertainty.}
We report the mean and sample standard deviation
($\mathrm{ddof}=1$) over three training seeds on one fixed split, and a
paired shot-level bootstrap with 2,000 replicates. Each replicate
resamples the test shots with replacement, separately within disruptive
and non-disruptive shots, from the test shots evaluated by every
compared method, and the same resampled shots are used for all methods
and seeds. Each metric is averaged over the seeds within a replicate,
and we report 95\% percentile intervals for the difference between
two methods. These intervals are conditional on the
fixed split and training seeds and do not describe variation across
data splits or devices; we therefore also show the value of every
training seed. The sensitivity of the fixed policy to the
calibration split is analyzed by resampling the calibration shots
1,000 times with the checkpoints and test data fixed
(\cref{app:supplementary-results}).

\subsection{Results}
\label{sec:results}

\subsubsection{Comparison of Training Targets}
\label{sec:results-main}
\label{sec:results-target}
\label{sec:results-calibration}

\begin{figure*}[t]
    \centering
    \includegraphics[width=0.9\textwidth]{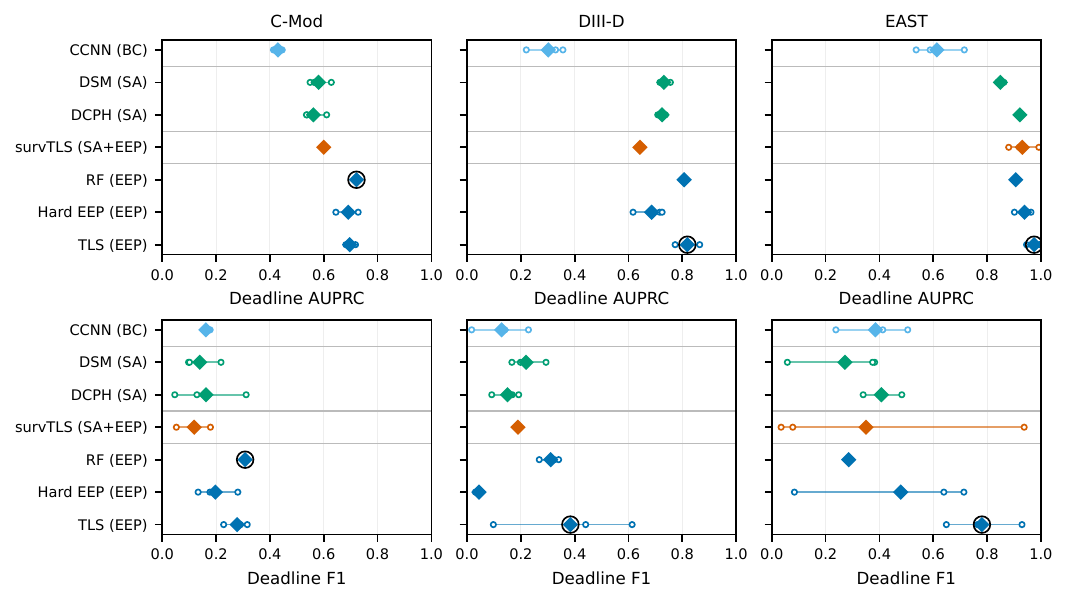}
    \caption{
        Deadline AUPRC (top) and validation-selected fixed-policy
        Deadline F1 (bottom) at $w=40\,\mathrm{ms}$. Open circles show
        the three training seeds, diamonds their mean, and lines the
        range across seeds. Colors indicate the training target; the
        circled point marks the largest mean on each device. Hard EEP
        is trained at $H=h_{\min}$ of the selected TLS bounds.
    }
    \label{fig:main-results}
\end{figure*}

\Cref{fig:main-results} compares the methods at the primary deadline
$w=40\,\mathrm{ms}$. Configurations are selected on the monitor split
with seed 0 and frozen for seeds 1 and 2. Because the fixed-policy
alarm metrics vary strongly across training seeds, the figure shows
every seed together with the mean. Numerical values and paired
differences are reported in
\cref{tab:main-results,tab:paired-differences}.

\begin{table*}[t]
    \centering
    \scriptsize
    \setlength{\tabcolsep}{1.5pt}
    \caption{
        Label-smoothing ablation at $w=40\,\mathrm{ms}$: TLS with the
        selected bounds $(h_{\min},h_{\max})$ and Hard EEP with
        $H=h_{\min}$ and $H=h_{\max}$, trained on the same cohort with
        the same encoder and optimization. Mean $\pm$ sample standard
        deviation over three training seeds, and TLS minus Hard EEP with
        95\% paired-bootstrap intervals; an asterisk marks intervals
        that exclude zero.
    }
    \label{tab:smoothing-ablation}
    \input{manuscript/tables/smoothing_ablation}
\end{table*}

\begin{figure*}[t]
    \centering
    \includegraphics[width=0.82\textwidth]{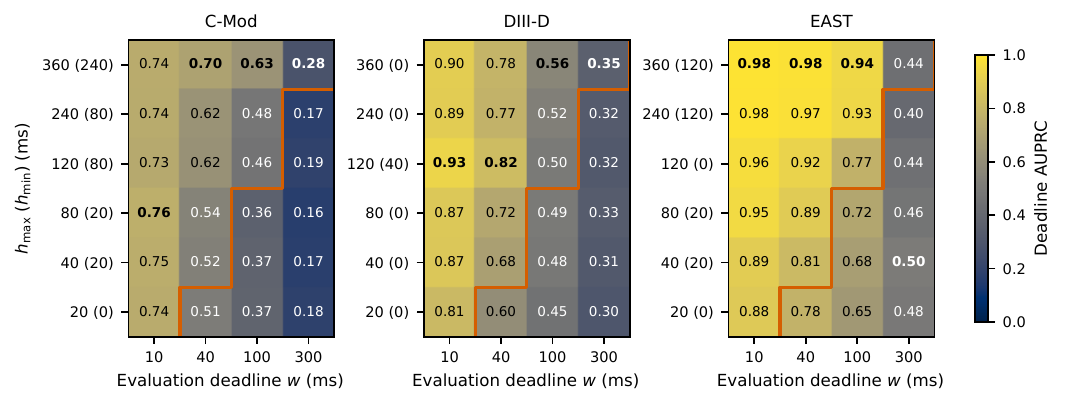}
    \caption{
        Deadline AUPRC of TLS for each upper bound $h_{\max}$ of the
        target and evaluation deadline $w$, averaged over three training
        seeds. Row labels give $h_{\max}$ and, in parentheses, the lower
        bound $h_{\min}$ selected for it. The orange line separates cells
        with $w\leq h_{\max}$ (left) from cells with $w>h_{\max}$
        (right); bold values mark the largest value in each column.
    }
    \label{fig:hmax-w-ablation}
\end{figure*}

\paragraph{EEP versus survival analysis.}
On every device, the best EEP model exceeds both DSM and DCPH in
Deadline AUPRC and Deadline F1 (\cref{tab:paired-differences}). The best
EEP model is TLS on DIII-D (Deadline AUPRC $0.819$, Deadline F1 $0.384$)
and EAST ($0.975$, $0.781$), and RF on C-Mod ($0.722$, $0.308$). TLS
exceeds RF on DIII-D in Deadline F1 by $+0.073$ ($[+0.025,\,+0.123]$)
without a resolved difference in Deadline AUPRC, and on EAST in both
metrics. On C-Mod, RF is slightly ahead of TLS in both metrics, but
neither difference is resolved.

survTLS, which restricts the survival objective to the horizon, does
not consistently outperform the full-distribution survival models: its
mean Deadline AUPRC is higher than those of DSM and DCPH on C-Mod and
EAST but lower on DIII-D, and its mean Deadline F1 exceeds both on none
of the devices. Because survTLS and DSM also differ in the
parameterization of the residual time, the target smoothing, and the
alarm score (\cref{app:alarm-scores}), this comparison does not isolate
horizon truncation, but it shows little benefit of truncation relative
to DSM.

\paragraph{Binary classification.}
The successful alarms of CCNN are raised much earlier than those of the
other methods, with a median warning time above $1\,\mathrm{s}$ on
DIII-D and EAST. A shot-level target does not tell the model when the
disruption occurs, so its scores often rise early in a discharge; such
early excursions also occur in non-disruptive shots, where they count as
false alarms.

\paragraph{Device dependence.}
Alarm performance remains limited on C-Mod and DIII-D for every method:
the largest mean Deadline F1 is $0.308$ on C-Mod and $0.384$ on DIII-D,
compared with $0.781$ on EAST. C-Mod is also the most difficult device
in ranking. When the deadline is shortened to $w=10\,\mathrm{ms}$,
however, every TLS configuration on C-Mod reaches a Deadline AUPRC of
$0.73$--$0.76$ (\cref{fig:hmax-w-ablation}), which is consistent with
precursors that develop shortly before the disruption.

\paragraph{Ranking versus calibrated alarms.}
The fixed-policy alarm metrics vary much more across seeds than the
ranking metric. The Deadline F1 of TLS on DIII-D is $0.098$, $0.441$,
and $0.613$ for the three seeds, whereas its Deadline AUPRC varies only
from $0.773$ to $0.865$; Hard EEP and survTLS on EAST show the same
pattern. The shot-level bootstrap intervals are conditional on the
trained seeds and do not include this variation. Threshold selection
contributes to it, because each seed uses a calibration split with
fewer than 1,200 shots (\cref{tab:partition-counts}). Resampling the
calibration shots with each checkpoint fixed moves the Deadline F1 of
TLS on DIII-D within $[0.288,\,0.526]$ and that of survTLS on EAST
within $[0.343,\,0.551]$ (seed-averaged 2.5th and 97.5th percentiles;
\cref{tab:calibration-summary}). The Recall--FAR curves
(\cref{fig:recall-far}) show that the ranking supports the fixed-policy
results at the primary budgets: on DIII-D and EAST, TLS reaches the
highest Deadline Recall near the primary FAR budget, whereas on C-Mod
the curves of TLS, RF, and Hard EEP overlap. Fixed-policy alarm metrics
should therefore be interpreted together with the threshold-free
ranking and with their variation across seeds and calibration data.

\subsubsection{Ablation: Label Smoothing}
\label{sec:results-smoothing}

The hard targets at $H=h_{\min}$ and $H=h_{\max}$ bound the TLS target
(\cref{eq:method-tls-bounds}), so comparing TLS with Hard EEP at both
bounds, with the cohort, encoder, and optimization fixed, isolates the
effect of the smoothing (\cref{tab:smoothing-ablation}).

TLS exceeds Hard EEP at both bounds in Deadline F1 on all three
devices, and every paired interval excludes zero: by $+0.340$ and
$+0.252$ on DIII-D, $+0.081$ and $+0.069$ on C-Mod, and $+0.302$ and
$+0.157$ on EAST, for $H=h_{\min}$ and $H=h_{\max}$, respectively. Hard
EEP at $H=h_{\max}$ has a higher mean Deadline F1 than at $H=h_{\min}$ on
every device: extending the positive region already helps the hard
target, but the graded target helps more.

\subsubsection{Ablation: Label Range and Evaluation Deadline}
\label{sec:results-horizon-ablation}

For each upper bound $h_{\max}$, the lower bound $h_{\min}$ is selected
on the monitor split at $w=40\,\mathrm{ms}$ with seed 0 and frozen for
all deadlines and seeds. \Cref{fig:hmax-w-ablation} reports the
Deadline AUPRC of these configurations for every evaluation deadline
$w$. The rows differ in $h_{\min}$ and, for some intervals taken from
earlier runs, in the training cohort (\cref{tab:tls-candidates}), so
the comparison across rows is descriptive.

The Deadline AUPRC decreases as the evaluation deadline $w$ grows, and
an upper bound $h_{\max}$ that covers the deadline mitigates this
decrease. This behavior is expected: an earlier warning is a harder
task, and $h_{\max}$ sets the range of residual times over which the
target carries information about the risk. At $w=300\,\mathrm{ms}$,
which exceeds every upper bound except the largest, the Deadline AUPRC
is at most $0.50$ on all devices. The target should therefore extend to
the lead time required by the intended control action.

\subsubsection{Ablation: Encoder Architecture}
\label{sec:results-encoder-ablation}

We fix the TLS objective, data partitions, training cohort, and
selected bounds, and remove either the recent-window branch (GRU-only)
or the history branch (MLP-only) (\cref{tab:tls-encoder-ablation}).

\begin{table}[t]
    \centering
    \scriptsize
    \setlength{\tabcolsep}{2.5pt}
    \caption{
        TLS encoder ablation: Deadline AUPRC and Deadline F1 at
        $w=40\,\mathrm{ms}$. Bold entries mark the largest mean within
        each device.
    }
    \label{tab:tls-encoder-ablation}
    \input{manuscript/tables/encoder_ablation}
\end{table}

The history branch is essential on every device: MLP-only yields the
lowest Deadline AUPRC and Deadline F1 on all three devices. The
contribution of the recent-window branch is device dependent: GRU+MLP
achieves the best Deadline AUPRC and Deadline F1 on DIII-D and C-Mod,
whereas GRU-only is best on EAST. A gating mechanism that can attenuate
the recent-window features when they are not useful may therefore be
beneficial.

\FloatBarrier

%% file: manuscript/tables/smoothing_ablation.tex
\begin{tabular}{llrrll}
\toprule
Device & Target (ms) & \makecell{Deadline\\AUPRC} & \makecell{Deadline\\F1} & \makecell{TLS $-$ Hard:\\$\Delta$ AUPRC [95\% CI]} & \makecell{TLS $-$ Hard:\\$\Delta$ F1 [95\% CI]} \\
\midrule
\multirow{3}{*}{C-Mod} & Hard $H=h_{\min}=240$ & $0.691\pm0.042$ & $0.197\pm0.076$ & $+0.005$ $[-0.013,\,+0.024]$ & $+0.081$ $[+0.044,\,+0.118]^{*}$ \\
 & Hard $H=h_{\max}=360$ & $0.717\pm0.016$ & $0.209\pm0.061$ & $-0.020$ $[-0.042,\,+0.002]$ & $+0.069$ $[+0.036,\,+0.104]^{*}$ \\
 & TLS $(240,\,360)$ & $0.696\pm0.019$ & $0.278\pm0.045$ &  &  \\
\midrule
\multirow{3}{*}{DIII-D} & Hard $H=h_{\min}=40$ & $0.686\pm0.060$ & $0.044\pm0.011$ & $+0.133$ $[+0.098,\,+0.162]^{*}$ & $+0.340$ $[+0.301,\,+0.377]^{*}$ \\
 & Hard $H=h_{\max}=120$ & $0.734\pm0.044$ & $0.132\pm0.097$ & $+0.085$ $[+0.049,\,+0.118]^{*}$ & $+0.252$ $[+0.213,\,+0.292]^{*}$ \\
 & TLS $(40,\,120)$ & $0.819\pm0.046$ & $0.384\pm0.262$ &  &  \\
\midrule
\multirow{3}{*}{EAST} & Hard $H=h_{\min}=120$ & $0.939\pm0.033$ & $0.479\pm0.344$ & $+0.036$ $[+0.028,\,+0.044]^{*}$ & $+0.302$ $[+0.283,\,+0.320]^{*}$ \\
 & Hard $H=h_{\max}=240$ & $0.947\pm0.036$ & $0.624\pm0.154$ & $+0.028$ $[+0.021,\,+0.035]^{*}$ & $+0.157$ $[+0.136,\,+0.177]^{*}$ \\
 & TLS $(120,\,240)$ & $0.975\pm0.025$ & $0.781\pm0.141$ &  &  \\
\bottomrule
\end{tabular}

%% file: manuscript/tables/encoder_ablation.tex
\begin{tabular}{llrr}
\toprule
Device & Encoder & AUPRC & F1 \\
\midrule
\multirow{3}{*}{C-Mod} & GRU+MLP & $\mathbf{0.696}\pm0.019$ & $\mathbf{0.278}\pm0.045$ \\
 & GRU-only & $0.670\pm0.030$ & $0.215\pm0.065$ \\
 & MLP-only & $0.618\pm0.014$ & $0.061\pm0.020$ \\
\midrule
\multirow{3}{*}{DIII-D} & GRU+MLP & $\mathbf{0.819}\pm0.046$ & $\mathbf{0.384}\pm0.262$ \\
 & GRU-only & $0.745\pm0.041$ & $0.066\pm0.019$ \\
 & MLP-only & $0.678\pm0.029$ & $0.025\pm0.044$ \\
\midrule
\multirow{3}{*}{EAST} & GRU+MLP & $0.975\pm0.025$ & $0.781\pm0.141$ \\
 & GRU-only & $\mathbf{0.994}\pm0.004$ & $\mathbf{0.941}\pm0.017$ \\
 & MLP-only & $0.960\pm0.005$ & $0.256\pm0.016$ \\
\bottomrule
\end{tabular}

%% file: manuscript/conclusion.tex
\section{Conclusion and Limitations}
\label{sec:conclusion}

This study asked which risk a disruption predictor should provide to an
alarm policy, and compared BC, full-distribution and horizon-truncated
SA, and fixed-horizon EEP on the DIII-D, Alcator C-Mod, and EAST
databases of DisruptionBench under a common causal protocol. On every
device, TLS outperformed DSM and DCPH in both Deadline AUPRC and Deadline
F1 at a $40\,\mathrm{ms}$ deadline. TLS performed best on DIII-D and
EAST, and a random forest trained on a hard horizon target performed
best on C-Mod. TLS also exceeded hard targets placed at either end of
its smoothing range in Deadline F1 on all three devices, so the graded
target itself, rather than the position of the label boundary, improves
the alarms. Truncating the survival objective at the horizon showed
little benefit relative to full-distribution SA. For the risk that
drives a disruption alarm, these results favor learning the risk within
an operational horizon directly over modeling the residual time over
its full range.

A fixed-horizon model provides little information beyond the upper
bound of its target, so the label range should cover the warning time
required by the intended action. Fixed-policy alarm metrics varied much more
across seeds than the ranking of the underlying risk, so alarm
thresholds should be evaluated together with that ranking and across
seeds. Alarm performance remained limited on C-Mod and DIII-D, which
suggests that the global plasma quantities used here capture only part
of the precursor information on these devices.

This study has several limitations. All results are within-device and
were obtained on one fixed split with three training seeds.
Configurations were selected with one seed, and TLS was selected from
more candidates than the other methods. The inputs are limited to the
global plasma quantities of DisruptionBench, CCNN was reimplemented
under the present protocol, and the evaluation is offline rather than in
a real-time PCS. Future work includes stronger representations,
multiple horizons for mitigation and avoidance, and cross-device
transfer of EEP models.

%% file: manuscript/appendix.tex

\begin{appendices}
\makeatletter
\@addtoreset{table}{section}
\@addtoreset{figure}{section}
\@addtoreset{equation}{section}
\makeatother
\section{Notation}
\label{app:notation}

\Cref{tab:notation} summarizes the notation used in the training and
evaluation formulations.

\begin{table*}[htbp]
    \centering
    \small
    \caption{Principal notation used in the training and evaluation formulations.}
    \label{tab:notation}
    \begin{tabular}{lll}
        \toprule
        Symbol & Meaning & Unit or setting \\
        \midrule
        $\Delta_x$
        & Diagnostic input-grid interval
        & $5\,\mathrm{ms}$
        \\
        $H$
        & Physical training horizon
        & $\mathrm{ms}$
        \\
        $\delta_h$
        & survTLS output-bin width
        & $1$ or $5\,\mathrm{ms}$
        \\
        $K_H=\lceil H/\delta_h\rceil$
        & Number of survTLS hazard bins within $H$
        & bins
        \\
        $w$
        & Required alarm lead time
        & $\mathrm{ms}$
        \\
        $\delta_i$
        & Event indicator for shot $i$
        & binary
        \\
        $\mathcal{I}_i$
        & Valid causal landmark set for shot $i$
        & landmarks
        \\
        $\rho_{\gamma}$
        & Exponential TLS smoothing profile
        & dimensionless
        \\
        $h_{\min}$, $h_{\max}$
        & Lower and upper bounds of the TLS smoothing range
        & $\mathrm{ms}$
        \\
        $q_{i,t}^{\mathrm{TLS}}$
        & TLS soft target
        & value in $[0,1]$
        \\
        $\widetilde{f}_{i,t}(k)$
        & survTLS smoothed event-time mass
        & probability mass
        \\
        $u_{i,t,k}$
        & survTLS soft at-risk weight
        & value in $[0,1]$
        \\
        $r_{i,t}^{(m)}(H)$
        & Alarm score for method $m$
        & dimensionless
        \\
        $v_i^{(w)}$
        & Threshold-free deadline score for shot $i$
        & dimensionless
        \\
        $\psi_{\alpha}^{\mathrm{exp}}$
        & Exponential alarm-priority function
        & dimensionless
        \\
        $\gamma$
        & TLS smoothing curvature
        & $0.025\,\mathrm{ms}^{-1}$
        \\
        $\alpha$
        & survTLS alarm-priority rate
        & $0.025\,\mathrm{ms}^{-1}$
        \\
        $\ell$
        & survTLS target length scale
        & dimensionless
        \\
        \bottomrule
    \end{tabular}
\end{table*}

\section{Detailed Experimental Configuration}
\label{app:experiment-details}

\subsection{Diagnostic Quantities}
\label{app:diagnostic-quantities}

\Cref{tab:diagnostic-quantities} lists the input channels. The nine
quantities in the upper block are common to the three databases. The
neural predictors on DIII-D and EAST additionally receive the three
device-specific signals in the lower block, so that each device has
twelve input channels. The C-Mod input instead contains three
machine-indicator channels, which are constant within the
within-device task and become zero after standardization. RF uses only
the nine common quantities. Definitions and units of the common
quantities follow \citet{spangher2025disruptionbench}. The
device-specific signals are not documented there; we list them by
their database column names and do not interpret them physically.

\begin{table*}[t]
    \centering
    \small
    \caption{Input channels. Upper block: quantities common to the three
        databases. Lower block: device-specific signals, given by their
        database names.}
    \label{tab:diagnostic-quantities}
    \begin{tabular}{@{}lp{0.6\textwidth}l@{}}
        \toprule
        Symbol & Physical quantity & Unit \\
        \midrule
        $\beta_p$
        & Plasma pressure normalized by poloidal magnetic pressure
        & dimensionless
        \\
        $\ell_i$
        & Normalized plasma internal inductance
        & dimensionless
        \\
        $q_{95}$
        & Safety factor at 95\% normalized flux
        & dimensionless
        \\
        $B_{n=1}$
        & $n=1$ component of the perturbed magnetic field
        & T
        \\
        $n/n_G$
        & Electron density normalized by the Greenwald limit
        & dimensionless
        \\
        $g_{\mathrm{lower}}$
        & Gap between the plasma and lower divertor
        & m
        \\
        $\kappa$
        & Plasma elongation
        & dimensionless
        \\
        $I_{p,\mathrm{error}}/I_{p,\mathrm{prog}}$
        & Plasma-current tracking error normalized by the programmed
          current
        & dimensionless
        \\
        $V_{\mathrm{loop}}$
        & Toroidal loop voltage
        & V
        \\
        \midrule
        \texttt{n1rms\_normalized}
        & DIII-D database column
        & --
        \\
        \texttt{radiated\_fraction}
        & DIII-D database column
        & --
        \\
        \texttt{zcur}
        & DIII-D database column
        & --
        \\
        \texttt{Mirnov\_std\_normalized}
        & EAST database column
        & --
        \\
        \texttt{rad\_input\_frac}
        & EAST database column
        & --
        \\
        \texttt{z\_error}
        & EAST database column
        & --
        \\
        \bottomrule
    \end{tabular}
\end{table*}

\begin{table*}[t]
    \centering
    \small
    \caption{Common optimization and reproducibility settings.}
    \label{tab:common-optimization-settings}
    \begin{tabular}{ll}
        \toprule
        Setting & Value \\
        \midrule
        Common time grid & $5\,\mathrm{ms}$ ($200\,\mathrm{Hz}$) \\
        Prepared-data split & Fixed across all runs \\
        Training seeds & 0, 1, and 2 \\
        Reported uncertainty & Sample standard deviation across training seeds
        ($\mathrm{ddof}=1$) \\
        Encoder & one-layer history GRU plus recent-window MLP \\
        GRU hidden dimension & 64 \\
        GRU output normalization & LayerNorm \\
        GRU truncated-BPTT length & 256 input steps \\
        Recent-window length & 10 samples ($50\,\mathrm{ms}$) \\
        Recent-window MLP & hidden dimension 64, depth 2 \\
        Encoder dropout & 0 \\
        Optimizer & AdamW \\
        Batch size & 50 \\
        Learning-rate schedule & cosine decay \\
        Warm-up & 10 epochs \\
        Gradient clipping & maximum norm 1.0 \\
        Early-stopping patience & 100 epochs \\
        Time / STFT features & disabled \\
        \bottomrule
    \end{tabular}
\end{table*}

\subsection{Shot Partitions and Eligibility}
\label{app:shot-partitions}

\begin{table}[t]
    \centering
    \small
    \caption{Fixed shot-level partition roles.}
    \label{tab:partition-roles}
    \begin{tabular}{lrp{2.6cm}}
        \toprule
        Partition & Fraction & Role \\
        \midrule
        Training & 60\% & Parameter optimization \\
        Calibration & 10\% & Alarm-threshold selection \\
        Monitoring & 10\% & Checkpoint and configuration selection \\
        Test & 20\% & Final held-out evaluation \\
        \bottomrule
    \end{tabular}
\end{table}

\paragraph{Data and split identification.}
The inputs are prepared from the database files \texttt{d3d\_for\_nn.mat},
\texttt{C\_Mod\_for\_nn.mat}, and \texttt{EAST\_for\_nn.mat} of the
DisruptionBench data archive \citep{spangher2025disruptionbench}.
Shots in which an input signal is never observed are removed
(808 DIII-D, 6 C-Mod, and 2,965 EAST shots), and the leading rows of a
shot are discarded until every signal has been observed, so that the
forward-fill never uses a later value. This leaves 7,800 DIII-D
(1,161 disruptive), 4,504 C-Mod (972), and 11,382 EAST (3,361) shots.
Shots are assigned to the four partitions by a random assignment
stratified by the shot outcome, with seed 42, which is identical for all
methods and seeds. Models were trained with revision \texttt{3b7239c}
of the project code (four seed-0 fits with revision \texttt{118a11d});
the follow-up TLS, Hard EEP, RF, and survTLS fits used revisions
\texttt{aeb50a1}, \texttt{bc97ea7}, \texttt{86170aa}, and
\texttt{d958342}, which add options for the target bounds, the
extended horizons, and the training cohort. The evaluation was computed with revisions
\texttt{214804b} and \texttt{78eb775} (follow-up). All survTLS
results were subsequently replaced by new fits with revision
\texttt{b0234ae}, which uses the at-risk weight in
\cref{eq:method-survtls-target-hazard}; their saved-prediction evaluation
used revision \texttt{dc30a05}. Other methods were retained unchanged.

\paragraph{Eligibility and partition sizes.}
A shot enters a partition only if its prepared sequence contains between
15 and 2,048 landmarks, which excludes some long EAST discharges, and, for a disruptive shot, at least one landmark
with $R_{i,t}\geq H_{\mathrm{elig}}$. For training, $H_{\mathrm{elig}}$
is the training horizon $H$; for evaluation, it is the deadline $w$.
\Cref{tab:partition-counts} gives the resulting partition sizes for
$H_{\mathrm{elig}}=40\,\mathrm{ms}$, which are also the test sets of the
evaluation at $w=40\,\mathrm{ms}$ and the common shot sets of the paired
bootstrap. For $H=120\,\mathrm{ms}$, between 15 (EAST) and 28 (DIII-D)
disruptive training shots and up to 9 disruptive test shots are
additionally excluded; the non-disruptive partitions do not change.

\begin{table}[t]
    \centering
    \scriptsize
    \setlength{\tabcolsep}{2.5pt}
    \caption{
        Partition sizes after preprocessing and eligibility at
        $H_{\mathrm{elig}}=40\,\mathrm{ms}$: disruptive / non-disruptive
        shots, and the total observation time of non-disruptive shots
        $D_0$ in hours.
    }
    \label{tab:partition-counts}
    \begin{tabular}{lccc}
        \toprule
        Partition & C-Mod & DIII-D & EAST \\
        \midrule
        Training    & 543 / 2,117 & 692 / 3,978 & 1,971 / 4,680 \\
        Calibration & 92 / 353    & 116 / 664   & 331 / 782 \\
        Monitoring  & 92 / 353    & 116 / 664   & 327 / 775 \\
        Test        & 184 / 707   & 225 / 1,324 & 657 / 1,561 \\
        \midrule
        Test $D_0$ (h) & 0.114 & 1.444 & 2.600 \\
        \bottomrule
    \end{tabular}
\end{table}

\paragraph{Normalization.}
Each input channel is standardized with the mean and standard deviation
estimated on all landmarks of the eligible training shots of the
corresponding run; the same statistics are applied to the calibration,
monitoring, and test partitions. Channels that are constant on the
training data, such as the C-Mod machine indicators, become zero.

\paragraph{Disruption time.}
The disruption time $T_i$ is taken from the time-to-disruption field of
the DisruptionBench databases \citep{spangher2025disruptionbench} and is
not redefined in this work. DisruptionBench does not specify a single
physical criterion for this time across the three devices; the
$40\,\mathrm{ms}$ deadline is motivated there by the minimum
activation time of the ITER disruption mitigation system.
For every disruptive shot, $t+R_{i,t}$ is constant to within
$1\,\mu\mathrm{s}$ over all retained landmarks, and the last retained
landmark lies at most one input-grid interval ($5\,\mathrm{ms}$) before
$T_i$ on DIII-D. On C-Mod and EAST, the last landmark of 633 and 5
disruptive shots, respectively, lies up to $10\,\mathrm{ms}$ and
$14\,\mathrm{ms}$ before $T_i$. The deadline $T_i-w$ is resolved on the
$5\,\mathrm{ms}$ input grid. Because inputs are forward-filled, the most
recent raw measurement at a landmark can be up to one native sampling
interval old ($5$, $25$, and $100\,\mathrm{ms}$ on C-Mod, DIII-D, and
EAST).

\subsection{Optimization and Reproducibility}
\label{app:optimization}

\Cref{tab:common-optimization-settings} lists the optimization and
encoder settings shared by all neural predictors except CCNN.

\subsection{Method-Specific Settings}
\label{app:method-settings}

\Cref{tab:method-specific-settings} lists the objective, the searched
settings, and the learning rates of each method, and
\cref{tab:tls-candidates} lists the candidate TLS smoothing intervals.
Each TLS candidate is trained on the cohort of shots that are eligible
for an eligibility horizon $H_{\mathrm{elig}}$
(\cref{app:shot-partitions}). The intervals in the last row of the
table were added in a follow-up search with a common cohort
($H_{\mathrm{elig}}=120\,\mathrm{ms}$). The interval
$(40,120)\,\mathrm{ms}$ appears twice, with cohorts of $40$ and
$120\,\mathrm{ms}$. For survTLS, the smoothing support is
$T_{\max}=120\,\mathrm{ms}$ for $H\leq80\,\mathrm{ms}$ and $T_{\max}=3H$
otherwise, and the alarm score of \cref{app:alarm-scores} is integrated
up to $H$.

\begin{table*}[t]
    \centering
    \scriptsize
    \caption{
        Method-specific settings. All neural predictors except CCNN
        share the encoder and the common optimization settings in
        \cref{tab:common-optimization-settings}.
    }
    \label{tab:method-specific-settings}
    \begin{tabular}{lp{3.6cm}p{7.2cm}l}
        \toprule
        Method & Objective & Settings & Learning rate \\
        \midrule
        DSM
        & Weibull mixture, lower bound of $\mathcal{L}_{\mathrm{SA}}$
        & $M\in\{1,3\}$; every valid landmark, loss averaged over time and
          then shots; 10,000 mixture-pretraining iterations at learning
          rate $10^{-2}$; at most 210 epochs
        & $10^{-3}$ \\
        DCPH
        & Cox partial likelihood
        & Risk sets from all valid landmarks in a mini-batch of shots;
          at most 210 epochs
        & $10^{-3}$ \\
        survTLS
        & $\mathcal{L}_{\mathrm{surv}}^{(H)}$ with $K_H=\lceil H/\delta_h\rceil$ hazards
        & $H\leq360\,\mathrm{ms}$; $\delta_h\in\{1,5\}\,\mathrm{ms}$; smoothing length scale 10;
          prevalence-aware output-bias initialization;
          $\alpha=0.025\,\mathrm{ms}^{-1}$ for the alarm-priority function; at most 210 epochs
        & $10^{-4}$, $10^{-3}$ \\
        Hard EEP
        & Masked BCE with $y^{(H)}_{i,t}$
        & $H=h_{\min}$ (main) or $h_{\max}$ of the selected TLS bounds, with
          the TLS training cohort; positive-class weight 1; at most 210 epochs
        & $10^{-3}$ \\
        TLS
        & Masked BCE with $q^{\mathrm{TLS}}_{i,t}$
        & $\gamma=0.025\,\mathrm{ms}^{-1}$; $(h_{\min},h_{\max})$ from
          \cref{tab:tls-candidates}; positive-class weight 1;
          at most 210 epochs
        & $10^{-3}$ \\
        RF
        & Random forest on $y^{(H)}_{i,t}$
        & $H\leq360\,\mathrm{ms}$; 63 causal window features of the nine common signals;
          500 trees; minimum leaf size 20; at most 200 uniformly
          sampled landmarks per training shot; no class weights
        & -- \\
        CCNN
        & Sequence-to-label BCE on $d_{i,t}$
        & Architecture of \citet{arnold2023continuous}; disruptive
          training sequences truncated at $T_i-H$; at most 210 epochs
        & $10^{-3}$ \\
        \bottomrule
    \end{tabular}
\end{table*}

\begin{table}[t]
    \centering
    \scriptsize
    \setlength{\tabcolsep}{3.0pt}
    \caption{
        Candidate TLS smoothing intervals $(h_{\min},h_{\max})$ in
        milliseconds and the eligibility horizon $H_{\mathrm{elig}}$ of
        their training cohort. The same 22 candidates are used on every
        device.
    }
    \label{tab:tls-candidates}
    \begin{tabular}{p{4.6cm}l}
        \toprule
        Intervals $(h_{\min},h_{\max})$ & $H_{\mathrm{elig}}$ \\
        \midrule
        $(20,120)$, $(40,120)$, $(80,120)$, $(120,360)$ & $h_{\min}$ \\
        $(0,20)$, $(0,40)$, $(0,80)$, $(0,120)$, $(40,80)$, $(40,120)$
        & $h_{\max}$ \\
        $(0,240)$, $(0,360)$, $(20,40)$, $(20,80)$,
        $(20,240)$, $(20,360)$, $(40,240)$, $(40,360)$, $(80,240)$,
        $(80,360)$, $(120,240)$, $(240,360)$ & $120$ \\
        \bottomrule
    \end{tabular}
\end{table}

\paragraph{Search spaces.}
DSM is trained on every valid landmark, with the loss averaged over
time within a shot and then over shots, because sampling one landmark
per shot in each epoch would give it far fewer training landmarks than
the EEP models; the sampled variant is reported in
\cref{tab:protocol-variants}. survTLS searches
$(\delta_h,\eta)\in\{(1\,\mathrm{ms},10^{-4}),(5\,\mathrm{ms},10^{-4}),(5\,\mathrm{ms},10^{-3})\}$,
because its original setting was the only one with a bin width finer
than the input grid and a reduced learning rate. RF and survTLS search
horizons up to $360\,\mathrm{ms}$, the largest upper bound of the TLS
candidates. Each method therefore has four candidate configurations per
device, except DSM (eight), RF (six), survTLS (eighteen), and TLS (22);
Hard EEP is not searched separately.

\paragraph{Horizon of DCPH.}
\label{app:dcph-horizon}
The partial likelihood of DCPH does not involve $H$, and for a given
checkpoint the horizon risk $1-\exp[-\Lambda_0(H)e^{g_{\Theta}}]$ is a
strictly increasing transform of the log-risk $g_{\Theta}$. On a fixed
set of shots and landmarks, $H$ therefore changes neither the deadline
ranking, nor the checkpoint selected by Deadline AUROC, nor the
operating points obtained from the quantile-based threshold candidates.
Results for DCPH can nevertheless differ across $H$ for two reasons.
First, a disruptive shot is eligible only if it has a landmark with
$R_{i,t}\geq H$, so a longer horizon removes a few short disruptive shots
from the training, calibration, and monitoring splits and thus changes
the training trajectory. Second, when $\Lambda_0(H)e^{g_{\Theta}}$ is
large, the horizon risk saturates numerically at one and ties the
highest-risk landmarks. On DIII-D, for example, $H=20$ and
$40\,\mathrm{ms}$ yield the same eligible shots and identical results on
every metric. The horizon search for DCPH therefore selects among
slightly different training cohorts rather than among prediction
horizons; we retain it so that every method is searched over a common
grid.

\paragraph{CCNN.}
\label{app:ccnn-differences}
In \citet{arnold2023continuous}, each shot is truncated
$40\,\mathrm{ms}$ before its end, and the AUC is computed from a single
score per shot, so that score excursions earlier in a non-disruptive
shot do not count as false alarms. In our protocol, a non-disruptive
shot is scored by the maximum over its whole observed trajectory, every
alarm episode contributes to the FAR, and the deadline metrics depend on
when an alarm is first raised. The original implementation also used a
learning rate of $2\times10^{-2}$, a fixed $40\,\mathrm{ms}$ truncation,
and random clipping of non-disruptive training shots, whereas we use the
common learning rate of $10^{-3}$ and select the truncation horizon from
the same grid as the survival models. CCNN is therefore a
reimplementation evaluated under the present protocol, not a
reproduction of the published results.

\subsection{survTLS Implementation}
\label{app:survtls-implementation}

Let $r_{i,t}=R_{i,t}/\delta_h$ be the residual time in bins and
$s_{i,t}=\max(r_{i,t}/\ell,\epsilon)$ the standard deviation of the
smoothing Gaussian in bins, with $\ell=10$ and $\epsilon=10^{-8}$, and
let $\Phi_{i,t}(x)=\Phi\bigl((x-r_{i,t})/s_{i,t}\bigr)$, where $\Phi$ is
the standard normal distribution function. For a disruptive shot, the
smoothed event-time mass over the $K=T_{\max}/\delta_h$ output bins is
$\widetilde{f}_{i,t}(k) = \Phi_{i,t}(e_k)-\Phi_{i,t}(e_{k-1})$,
for $k=1,\dots,K$, with bin edges $e_0=-\infty$, $e_k=k+\tfrac{1}{2}$ for $1\le k<K$, and
$e_K=+\infty$. The first bin therefore absorbs the Gaussian mass at negative and
sub-bin residual times, and the last bin absorbs the mass beyond
$T_{\max}$; the masses sum to one without further normalization. For
non-disruptive shots, $\widetilde{f}_{i,t}(k)=0$ for every $k$. Target
hazards are clipped to $[0,1]$, with the denominator of
\cref{eq:method-survtls-target-hazard} bounded below by $\epsilon$. The
loss in \cref{eq:method-lsurv} uses the first $K_H$ of the $K$ bins, and
the target masses are not renormalized after this truncation. The
output bias of each hazard logit is initialized to the logit of the mean
target hazard of that bin over the training landmarks. The values of
$\delta_h$, $H$, and $T_{\max}$ are listed in
\cref{tab:method-specific-settings,tab:tls-candidates}.

\subsection{Checkpoint Selection and Threshold Calibration}
\label{app:checkpoint-selection}

For DSM, TLS, and survTLS, the within-run checkpoint is selected by
monitor-split deadline AUROC at $w=40\,\mathrm{ms}$. For seed 0, the
configuration (training horizon, TLS smoothing interval, DSM mixture
count, and survTLS bin width and learning rate) is then selected by
monitor-split
deadline F1 after threshold calibration on the calibration split. These
configuration choices are frozen for seeds 1 and 2. Test data are not
used for checkpoint, configuration, or alarm-policy selection.

Validation calibration evaluates at most 512 automatically generated
threshold candidates. The alarm configuration is fixed to one-sample
smoothing, one required consecutive threshold crossing, zero hysteresis,
and zero cooldown; under this configuration it reduces to the rule in
\cref{app:evaluation-alarm-events}. The low, middle, and high budgets target
approximately $0.5\%$, $1\%$, and $5\%$ normal-shot FPR, respectively;
the middle budget is primary. Normal-shot FPR is the fraction of
non-disruptive shots with at least one alarm. For $N_0$ non-disruptive
shots with total normal exposure $D_0$, target $p_{\mathrm{FPR}}$ is
converted to the nominal event-based ceiling
$C(p_{\mathrm{FPR}})
=p_{\mathrm{FPR}}N_0(24\,\mathrm{h})/D_0$. Because FAR counts
repeated alarm events, this conversion is an approximate FPR
interpretation. The resulting low/middle/high ceilings are
$(110,220,1100)$ for DIII-D,
$(650,1300,6500)$ for C-Mod, and $(63,126,630)$ for EAST, in alarms per
24 normal observation hours. Test FAR is reported without clipping or
post-hoc threshold adjustment. These monitor- and calibration-based
selections define the validation-selected fixed-policy evaluation.


\section{Detailed Evaluation Definitions}
\label{app:evaluation-definitions}

\subsection{Alarm Scores}
\label{app:alarm-scores}

For DSM, the score at landmark $t$ is the cumulative event probability
within the training horizon:
$r_{i,t}^{\mathrm{DSM}}(H) = \int_{0}^{H} f_{\Theta} \left(\tau\mid\mathbf{z}_{i,t}\right) \,\mathrm{d}\tau$.
DCPH uses the corresponding horizon risk
$1-S_{\Theta}(H\mid\mathbf{z}_{i,t})$ implied by its proportional-hazards
model. For TLS and Hard EEP, the score is the model output, which is a
smoothed horizon-aware risk score for TLS and an estimate of the
fixed-horizon event probability for Hard EEP:
$r_{i,t}^{\mathrm{TLS}}(H) = p_{\theta}^{(H)}(\mathbf{z}_{i,t})$.
For survTLS, the horizon-resolved hazards defined in
\cref{sec:methods-lsurv} first give the cumulative event probability
\begin{equation}
    F_{\theta}
    \left(k\mid\mathbf{z}_{i,t}\right)
    =
    1-
    \prod_{j=1}^{k}
    \left[
        1-
        \lambda_{\theta}
        \left(j\mid\mathbf{z}_{i,t}\right)
    \right],
    \label{eq:experiment-survtls-cumulative-risk}
\end{equation}
and the alarm score aggregates these horizon-specific risks after
weighting them by imminence:
\begin{equation}
    r_{i,t}^{\mathrm{survTLS}}(H)
    =
    \max_{1\leq k\leq K_H}
    \psi_{\alpha}^{\mathrm{exp}}(k\delta_h)
    F_{\theta}
    \left(k\mid\mathbf{z}_{i,t}\right).
    \label{eq:experiment-survtls-alarm-score}
\end{equation}
Here, $\psi_{\alpha}^{\mathrm{exp}}$ is the exponential alarm-priority
function
\begin{equation}
    \psi_{\alpha}^{\mathrm{exp}}(\tau)
    =
    \begin{cases}
        \dfrac{e^{-\alpha\tau}-e^{-\alpha H}}{1-e^{-\alpha H}},
        & 0<\tau\leq H,
        \\[2mm]
        0,
        & \tau>H,
    \end{cases}
    \label{eq:experiment-survtls-priority}
\end{equation}
which decreases from approximately one at short offsets to zero at the
horizon. The parameter $\alpha>0$, in $\mathrm{ms}^{-1}$, controls how
strongly shorter offsets are prioritized; we use
$\alpha=0.025\,\mathrm{ms}^{-1}$ ($25\,\mathrm{s}^{-1}$).
RF and CCNN use their predicted class probabilities.

\subsection{Alarm Events and the Operational Deadline}
\label{app:evaluation-alarm-events}

For shot $i$, let $\mathcal{I}_i$ be the set of valid causal landmarks,
let $\delta_i\in\{0,1\}$ indicate whether a disruption is observed, and
let $T_i$ be the disruption time when $\delta_i=1$. Method $m$ produces
an alarm score $r_{i,t}^{(m)}(H)$ at landmark $t$ for training horizon
$H$.
Let $\phi$ denote the causal score averaging, persistence, and cooldown
configuration selected on validation. The pointwise threshold state is
$b_{i,t}(\eta) = \mathbb{I}\!\left\{r_{i,t}^{(m)}(H)\geq\eta\right\}$,
and the fixed causal alarm operator $\Phi_{\phi}$ emits
$\mathcal{A}_i(\eta,\phi) = \Phi_{\phi}\!\left( (r_{i,t}^{(m)}(H))_{t\in\mathcal{I}_i},\eta \right)$.
\label{eq:evaluation-alarm-events}
Each emitted time is counted as one alarm event; a sustained active state
is not counted once per landmark.

For a required lead time $w>0$, measured in milliseconds and defined
independently of the training horizon $H$, the last eligible landmark
of a disruptive shot is
$t_i^{\mathrm{dl}}(w) = \max\left\{ t\in\mathcal{I}_i: t\leq T_i-w \right\}$.
The shot is detected when at least one alarm is emitted no later than
this deadline,
\begin{equation}
    z_i(\eta,\phi;w)
    =
    \mathbb{I}\!\left\{
        \min\mathcal{A}_i(\eta,\phi)\leq t_i^{\mathrm{dl}}(w)
    \right\},
    \label{eq:evaluation-deadline-hit}
\end{equation}
with $\min\emptyset=\infty$.
\label{eq:evaluation-event-detection}
There is no lower time bound: an alarm earlier than $T_i-w$ remains a
successful detection. An alarm after the deadline is late and does not
detect the shot.

\subsection{Deadline Recall, Alarm Precision, and F1}
\label{app:evaluation-deadline-prf}

Only one true alarm is credited for each detected disruptive shot. Every
remaining alarm event---including alarms on normal shots and late or
duplicate alarms on disruptive shots---is false. Thus,
\begin{align}
    N_{\mathrm{TP}}(\eta,\phi;w)
    &=
    \sum_{i:\delta_i=1} z_i(\eta,\phi;w),
    \\
    N_{\mathrm{FP}}(\eta,\phi;w)
    &=
    \sum_i \left|\mathcal{A}_i(\eta,\phi)\right|
    -N_{\mathrm{TP}}(\eta,\phi;w),
    \\
    N_{\mathrm{FN}}(\eta,\phi;w)
    &=
    \sum_i \delta_i-N_{\mathrm{TP}}(\eta,\phi;w).
    \label{eq:evaluation-alarm-counts}
\end{align}
The fixed-policy metrics are
\begin{align}
    \operatorname{Rec}_{w}(\eta,\phi)
    &=
    \frac{N_{\mathrm{TP}}}
         {N_{\mathrm{TP}}+N_{\mathrm{FN}}},
    \\
    \operatorname{Prec}_{\mathrm{alarm},w}(\eta,\phi)
    &=
    \frac{N_{\mathrm{TP}}}
         {N_{\mathrm{TP}}+N_{\mathrm{FP}}},
    \\
    F_{1,w}(\eta,\phi)
    &=
    \frac{2N_{\mathrm{TP}}}
         {2N_{\mathrm{TP}}+N_{\mathrm{FP}}+N_{\mathrm{FN}}}.
    \label{eq:evaluation-deadline-prf}
\end{align}
\label{eq:evaluation-event-alarm-pr}
When a denominator is zero, the corresponding precision or F1 is
reported as undefined in a results table and treated as zero for model
selection. The main experiment sets $w=40\,\mathrm{ms}$.

\subsection{FAR}
\label{app:evaluation-far}

FAR is computed only from non-disruptive shots. Let $D_i$ be the valid
observed duration of normal shot $i$, including its final sampling
interval. The total normal observation time is
$D_0=\sum_{i:\delta_i=0}D_i$, and the number of false alarms on normal
shots is $N_{\mathrm{FA},0}(\eta,\phi)=\sum_{i:\delta_i=0}|\mathcal{A}_i(\eta,\phi)|$.
The normalized FAR is
\begin{equation}
    \operatorname{FAR}_{24}(\eta,\phi)
    =
    \frac{N_{\mathrm{FA},0}(\eta,\phi)}{D_0}
    \left(24\,\mathrm{h}\right).
    \label{eq:evaluation-far24}
\end{equation}
Repeated alarm events on the same normal shot all contribute to FAR.

\subsection{Threshold-Free Deadline AUROC and AUPRC}
\label{app:evaluation-deadline-discrimination}

For a disruptive shot, the threshold-free deadline score is the largest
risk available by the deadline; for a non-disruptive shot, it is the
largest risk over the observed record:
\begin{equation}
    v_i^{(w)}
    =
    \begin{cases}
        \displaystyle
        \max_{t\in\mathcal{I}_i:\,t\leq t_i^{\mathrm{dl}}(w)}
        r_{i,t}^{(m)}(H),
        & \delta_i=1,
        \\
        \displaystyle
        \max_{t\in\mathcal{I}_i}r_{i,t}^{(m)}(H),
        & \delta_i=0.
    \end{cases}
    \label{eq:evaluation-warning-score}
\end{equation}
A disruptive shot with no eligible landmark receives a score below the
valid score range. For a threshold $\eta$, define the corresponding
shot-level prediction by
$\widehat{d}_i(\eta;w) = \mathbb{I}\!\left\{v_i^{(w)}\geq\eta\right\}$.
Sweeping $\eta$ over the score range gives the shot-level true-positive
rate (TPR), FPR, and precision
\begin{align}
    \operatorname{TPR}_{w}(\eta)
    =\operatorname{Rec}_{w}^{\mathrm{shot}}(\eta)
    &=
    \frac{
        \sum_i \delta_i\widehat{d}_i(\eta;w)
    }{
        \sum_i \delta_i
    },
    \\
    \operatorname{FPR}_{w}(\eta)
    &=
    \frac{
        \sum_i (1-\delta_i)\widehat{d}_i(\eta;w)
    }{
        \sum_i (1-\delta_i)
    },
    \\
    \operatorname{Prec}_{w}^{\mathrm{shot}}(\eta)
    &=
    \frac{
        \sum_i \delta_i\widehat{d}_i(\eta;w)
    }{
        \sum_i \widehat{d}_i(\eta;w)
    }.
    \label{eq:evaluation-deadline-shot-rates}
\end{align}
This shot-level precision is distinct from the alarm-event precision in
\cref{eq:evaluation-deadline-prf}. Let
$\operatorname{ROC}_{w}(\xi)$ denote the ROC curve, expressed as TPR at
FPR $\xi$ along the threshold sweep, and let
$\operatorname{PR}_{w}(\nu)$ denote the PR curve, expressed as
shot-level precision at recall $\nu$. Deadline AUROC and AUPRC are
\begin{align}
    \operatorname{AUROC}_{\mathrm{deadline}}(w)
    &=
    \int_{0}^{1}\operatorname{ROC}_{w}(\xi)\,\mathrm{d}\xi,
    \label{eq:evaluation-warning-auroc}
    \\
    \operatorname{AUPRC}_{\mathrm{deadline}}(w)
    &=
    \int_{0}^{1}\operatorname{PR}_{w}(\nu)\,\mathrm{d}\nu.
    \label{eq:evaluation-warning-auprc}
\end{align}
For a finite evaluation set, the integrals are evaluated from the
empirical curves obtained by sweeping the distinct values of
$v_i^{(w)}$. These quantities depend only on the ranking of the deadline
scores and are independent of the deployed alarm threshold.

\subsection{Validation Calibration and Horizon Selection}
\label{app:evaluation-calibration-selection}

Let $\mathcal{G}^{\mathrm{cal}}_{m,H}$ be the finite set of candidate
threshold and alarm-post-processing configurations for method $m$ and
training horizon $H$, and let $C$ be the device-specific FAR ceiling
corresponding to a normal-shot FPR budget. The calibration policy is
\begin{equation}
    \begin{aligned}
    (\eta^{\star}_{m,H,C},\phi^{\star}_{m,H,C})
    &=
    \arg\max_{(\eta,\phi)\in\mathcal{G}^{\mathrm{cal}}_{m,H}}
        \operatorname{Rec}^{\mathrm{cal}}_{w}(\eta,\phi)
    \\
    &\text{subject to}\quad
        \operatorname{FAR}^{\mathrm{cal}}_{24}(\eta,\phi)\leq C.
    \end{aligned}
    \label{eq:evaluation-calibration-objective}
\end{equation}
Ties are resolved by higher alarm precision, lower FAR, and then a higher
threshold. If no candidate satisfies the budget, the minimum-FAR
candidate is retained. The calibrated policy is applied unchanged to the
disjoint monitor split. Let $\mathcal{H}_m$ be the set of candidate
configurations of method $m$ (\cref{sec:experiment-optimization}), indexed
here by $H$; for TLS, $H$ stands for the smoothing interval and its
cohort. For the primary middle budget, the selected configuration is
\begin{equation}
    \begin{aligned}
    H^{\star}_{m}
    =
    \arg\max_{H\in\mathcal{H}_m}
    F^{\mathrm{mon}}_{1,w}
    \!\left(
        \eta^{\star}_{m,H,C_{\mathrm{mid}}},
        \phi^{\star}_{m,H,C_{\mathrm{mid}}}
    \right).
    \end{aligned}
    \label{eq:evaluation-horizon-selection}
\end{equation}
with remaining ties resolved by monitor recall, alarm precision, lower
FAR, shorter horizon, and fewer mixture components. The selection uses
seed 0 and is frozen for seeds 1 and 2. Test data do not enter either selection.

\subsection{Recall--FAR Curves}
\label{app:evaluation-test-far-constrained}

For the validation-selected checkpoint, configuration, and
post-processing $\phi^{\star}_m$ of method $m$, the threshold is swept
over every critical level of the test scores. Each threshold $\eta$
gives one point
$\left( \operatorname{FAR}^{\mathrm{test}}_{24,m}(\eta,\phi^{\star}_m),\; \operatorname{Rec}^{\mathrm{test}}_{w,m}(\eta,\phi^{\star}_m) \right)$
of the Recall--FAR curve. For averaging over seeds, each curve is
resampled on a common logarithmic FAR grid, taking the largest recall
whose FAR does not exceed each grid value. Because the threshold is
swept with test labels, the curves describe the test score distribution
and are not a deployable threshold-selection procedure. The curves for
the selected configurations are shown in \cref{fig:recall-far}.

\section{Supplementary Results}
\label{app:supplementary-results}

\begin{figure*}[!tp]
    \centering
    \includegraphics[width=\textwidth]{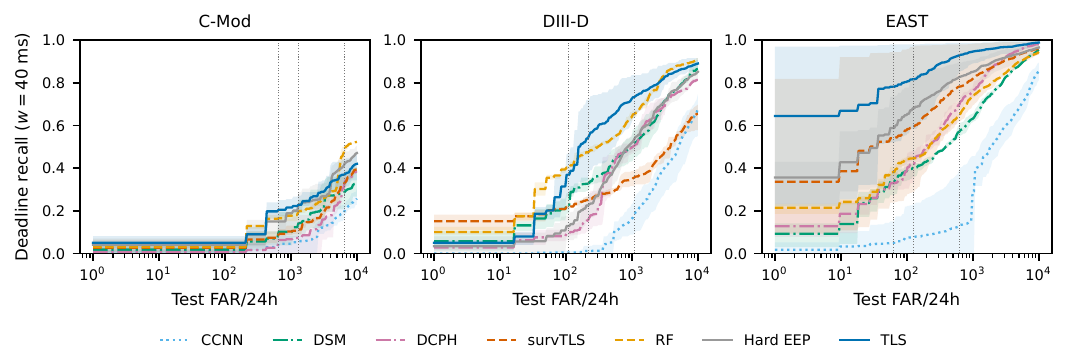}
    \caption{
        Deadline Recall versus realized test FAR/24h at
        $w=40\,\mathrm{ms}$, obtained by sweeping the threshold on the
        test data with the validation-selected post-processing fixed.
        Lines show the mean over three training seeds, and shaded bands
        the range. Dotted vertical lines mark the low, middle (primary),
        and high FAR budgets. The curves use test labels and are
        descriptive.
    }
    \label{fig:recall-far}
\end{figure*}

\Cref{tab:main-results} gives the numerical values underlying
\cref{fig:main-results}, together with Deadline AUROC and the
components of Deadline F1. \Cref{tab:paired-differences} gives the
paired differences between TLS and all other main methods, including
Hard EEP at both selected TLS bounds.
\Cref{fig:recall-far} shows the Recall--FAR curves, and
\cref{tab:calibration-summary} summarizes the sensitivity of the fixed
policy to the calibration split. \Cref{tab:multi-deadline} reports the
ranking of every method at the deadlines of the horizon study.
\Cref{tab:protocol-variants} compares the DSM and survTLS protocol
variants, and \cref{tab:disruptionbench-metrics} reports the DisruptionBench
metrics for comparison with previous benchmark results.

\begin{table*}[!tp]
    \centering
    \scriptsize
    \setlength{\tabcolsep}{1.0pt}
    \renewcommand{\arraystretch}{1.15}
    \caption{
        Results at the $w=40\,\mathrm{ms}$ deadline (mean $\pm$ sample
        standard deviation over three training seeds; the values of the
        individual seeds are shown in \cref{fig:main-results}). Deadline
        AUPRC and AUROC are threshold-free; the remaining metrics use the
        validation-selected fixed policy. The configuration column gives
        the seed-0 choice of the training horizon $H$ (ms), with the
        number of mixture components for DSM and the smoothing interval
        $(h_{\min},h_{\max})$ (ms) for TLS. Bold entries mark the largest
        mean Deadline AUPRC and Deadline F1 within each device.
    }
    \label{tab:main-results}
    \input{manuscript/tables/main_results}
\end{table*}

\begin{table*}[!tp]
    \centering
    \scriptsize
    \setlength{\tabcolsep}{2.5pt}
    \caption{
        Differences between TLS and each comparator (TLS minus
        comparator) at $w=40\,\mathrm{ms}$, with 95\% paired-bootstrap
        intervals from 2,000 stratified test-shot resamples, averaging
        the three fixed training seeds. These intervals condition on
        the selected configurations and exclude training-seed and
        selection uncertainty; no multiplicity adjustment is applied.
        An asterisk marks intervals that exclude zero.
    }
    \label{tab:paired-differences}
    \input{manuscript/tables/paired_differences}
\end{table*}

\begin{table}[!htbp]
    \centering
    \scriptsize
    \setlength{\tabcolsep}{3.0pt}
    \caption{
        Test Deadline F1 when the calibration shots are resampled 1,000
        times with the checkpoint and test data fixed. Median and
        2.5th--97.5th percentiles, averaged over three training seeds.
    }
    \label{tab:calibration-summary}
    \input{manuscript/tables/calibration_summary}
\end{table}

\begin{table}[!htbp]
    \centering
    \scriptsize
    \setlength{\tabcolsep}{1.5pt}
    \caption{
        Deadline AUPRC of the selected configurations at different
        evaluation deadlines $w$ (ms), averaged over three training
        seeds.
    }
    \label{tab:multi-deadline}
    \input{manuscript/tables/multi_deadline}
\end{table}

\begin{table}[!htbp]
    \centering
    \scriptsize
    \setlength{\tabcolsep}{1.5pt}
    \caption{
        Protocol variants. DSM with one sampled landmark per shot per
        epoch versus every landmark (main), and survTLS with its original
        setting ($\delta_h=1\,\mathrm{ms}$, $\eta=10^{-4}$) versus the
        reselected configuration (main): $(H,\delta_h,\eta)$ is
        $(40\,\mathrm{ms},1\,\mathrm{ms},10^{-4})$ on DIII-D,
        $(80\,\mathrm{ms},5\,\mathrm{ms},10^{-4})$ on C-Mod, and
        $(360\,\mathrm{ms},5\,\mathrm{ms},10^{-3})$ on EAST. Both survTLS
        variants use the corrected at-risk loss weight. Deadline AUPRC
        and Deadline F1 at $w=40\,\mathrm{ms}$, averaged over three
        training seeds.
    }
    \label{tab:protocol-variants}
    \input{manuscript/tables/protocol_variants}
\end{table}

\begin{table}[!htbp]
    \centering
    \scriptsize
    \setlength{\tabcolsep}{2.0pt}
    \caption{
        DisruptionBench metrics at the validation-selected threshold
        (equal high and low thresholds, one-sample hysteresis,
        $t_{\mathrm{useful}}=40\,\mathrm{ms}$), averaged over three
        training seeds.
    }
    \label{tab:disruptionbench-metrics}
    \input{manuscript/tables/disruptionbench_metrics}
\end{table}

\FloatBarrier
\end{appendices}

%% file: manuscript/tables/main_results.tex
\begin{tabular}{llcrrrrrr}
\toprule
Device & Method & Config. & \makecell{Deadline\\AUPRC} & \makecell{Deadline\\AUROC} & \makecell{Deadline\\Recall} & \makecell{Alarm\\precision} & \makecell{Deadline\\F1} & \makecell{Test\\FAR/24h} \\
\midrule
\multirow{7}{*}{C-Mod} & CCNN & $20$ & $0.430\pm0.016$ & $0.724\pm0.001$ & $0.094\pm0.013$ & $0.617\pm0.113$ & $0.162\pm0.014$ & $2{,}112\pm1{,}285$ \\
 & DSM & $80,\,3$ & $0.581\pm0.042$ & $0.808\pm0.009$ & $0.098\pm0.061$ & $0.268\pm0.042$ & $0.139\pm0.069$ & $1{,}056\pm634$ \\
 & DCPH & $120$ & $0.562\pm0.043$ & $0.803\pm0.007$ & $0.127\pm0.121$ & $0.254\pm0.118$ & $0.162\pm0.136$ & $2{,}887\pm2{,}520$ \\
 & survTLS & $80$ & $0.600\pm0.006$ & $0.817\pm0.007$ & $0.103\pm0.061$ & $0.142\pm0.062$ & $0.118\pm0.063$ & $1{,}267\pm762$ \\
 & RF & $240$ & $\mathbf{0.722}\pm0.003$ & $0.884\pm0.000$ & $0.293\pm0.009$ & $0.324\pm0.003$ & $\mathbf{0.308}\pm0.006$ & $4{,}647\pm1{,}118$ \\
 & Hard EEP & $240$ & $0.691\pm0.042$ & $0.866\pm0.022$ & $0.138\pm0.071$ & $0.386\pm0.017$ & $0.197\pm0.076$ & $775\pm976$ \\
 & TLS & $(240,\,360)$ & $0.696\pm0.019$ & $0.871\pm0.017$ & $0.228\pm0.039$ & $0.356\pm0.052$ & $0.278\pm0.045$ & $1{,}549\pm800$ \\
\midrule
\multirow{7}{*}{DIII-D} & CCNN & $120$ & $0.302\pm0.072$ & $0.703\pm0.071$ & $0.081\pm0.071$ & $0.376\pm0.112$ & $0.129\pm0.106$ & $338\pm242$ \\
 & DSM & $120,\,1$ & $0.731\pm0.021$ & $0.904\pm0.003$ & $0.188\pm0.091$ & $0.294\pm0.042$ & $0.220\pm0.066$ & $94\pm75$ \\
 & DCPH & $80$ & $0.725\pm0.014$ & $0.908\pm0.004$ & $0.136\pm0.064$ & $0.184\pm0.022$ & $0.151\pm0.052$ & $216\pm175$ \\
 & survTLS & $40$ & $0.642\pm0.007$ & $0.849\pm0.003$ & $0.138\pm0.012$ & $0.309\pm0.046$ & $0.189\pm0.003$ & $6\pm10$ \\
 & RF & $120$ & $0.807\pm0.005$ & $0.933\pm0.002$ & $0.274\pm0.076$ & $0.391\pm0.075$ & $0.311\pm0.038$ & $39\pm10$ \\
 & Hard EEP & $40$ & $0.686\pm0.060$ & $0.903\pm0.015$ & $0.039\pm0.010$ & $0.053\pm0.013$ & $0.044\pm0.011$ & $44\pm63$ \\
 & TLS & $(40,\,120)$ & $\mathbf{0.819}\pm0.046$ & $0.935\pm0.018$ & $0.409\pm0.295$ & $0.368\pm0.229$ & $\mathbf{0.384}\pm0.262$ & $238\pm277$ \\
\midrule
\multirow{7}{*}{EAST} & CCNN & $120$ & $0.613\pm0.092$ & $0.764\pm0.053$ & $0.290\pm0.141$ & $0.715\pm0.170$ & $0.385\pm0.136$ & $686\pm557$ \\
 & DSM & $80,\,3$ & $0.850\pm0.012$ & $0.915\pm0.007$ & $0.222\pm0.165$ & $0.415\pm0.102$ & $0.272\pm0.185$ & $34\pm23$ \\
 & DCPH & $120$ & $0.922\pm0.003$ & $0.963\pm0.003$ & $0.331\pm0.094$ & $0.546\pm0.009$ & $0.407\pm0.072$ & $55\pm37$ \\
 & survTLS & $360$ & $0.931\pm0.057$ & $0.964\pm0.027$ & $0.340\pm0.530$ & $0.413\pm0.447$ & $0.350\pm0.509$ & $28\pm48$ \\
 & RF & $80$ & $0.907\pm0.001$ & $0.955\pm0.000$ & $0.217\pm0.002$ & $0.418\pm0.012$ & $0.286\pm0.004$ & $3\pm5$ \\
 & Hard EEP & $120$ & $0.939\pm0.033$ & $0.963\pm0.017$ & $0.437\pm0.343$ & $0.661\pm0.118$ & $0.479\pm0.344$ & $25\pm23$ \\
 & TLS & $(120,\,240)$ & $\mathbf{0.975}\pm0.025$ & $0.984\pm0.013$ & $0.756\pm0.201$ & $0.815\pm0.079$ & $\mathbf{0.781}\pm0.141$ & $52\pm56$ \\
\bottomrule
\end{tabular}

%% file: manuscript/tables/paired_differences.tex
\begin{tabular}{llll}
\toprule
Device & Comparator & $\Delta$ Deadline AUPRC [95\% CI] & $\Delta$ Deadline F1 [95\% CI] \\
\midrule
\multirow{7}{*}{C-Mod} & CCNN & $+0.267$ $[+0.209,\,+0.319]^{*}$ & $+0.116$ $[+0.047,\,+0.179]^{*}$ \\
 & DSM & $+0.116$ $[+0.070,\,+0.161]^{*}$ & $+0.139$ $[+0.096,\,+0.179]^{*}$ \\
 & DCPH & $+0.135$ $[+0.089,\,+0.177]^{*}$ & $+0.116$ $[+0.075,\,+0.157]^{*}$ \\
 & survTLS & $+0.096$ $[+0.043,\,+0.154]^{*}$ & $+0.160$ $[+0.105,\,+0.212]^{*}$ \\
 & RF & $-0.025$ $[-0.074,\,+0.028]$ & $-0.030$ $[-0.087,\,+0.023]$ \\
 & Hard EEP ($H=h_{\min}$) & $+0.005$ $[-0.013,\,+0.024]$ & $+0.081$ $[+0.044,\,+0.118]^{*}$ \\
 & Hard EEP ($H=h_{\max}$) & $-0.020$ $[-0.042,\,+0.002]$ & $+0.069$ $[+0.036,\,+0.104]^{*}$ \\
\midrule
\multirow{7}{*}{DIII-D} & CCNN & $+0.517$ $[+0.463,\,+0.563]^{*}$ & $+0.256$ $[+0.205,\,+0.305]^{*}$ \\
 & DSM & $+0.087$ $[+0.042,\,+0.131]^{*}$ & $+0.165$ $[+0.123,\,+0.209]^{*}$ \\
 & DCPH & $+0.094$ $[+0.048,\,+0.137]^{*}$ & $+0.233$ $[+0.187,\,+0.277]^{*}$ \\
 & survTLS & $+0.176$ $[+0.122,\,+0.232]^{*}$ & $+0.195$ $[+0.147,\,+0.247]^{*}$ \\
 & RF & $+0.012$ $[-0.028,\,+0.053]$ & $+0.073$ $[+0.025,\,+0.123]^{*}$ \\
 & Hard EEP ($H=h_{\min}$) & $+0.133$ $[+0.098,\,+0.162]^{*}$ & $+0.340$ $[+0.301,\,+0.377]^{*}$ \\
 & Hard EEP ($H=h_{\max}$) & $+0.085$ $[+0.049,\,+0.118]^{*}$ & $+0.252$ $[+0.213,\,+0.292]^{*}$ \\
\midrule
\multirow{7}{*}{EAST} & CCNN & $+0.361$ $[+0.335,\,+0.385]^{*}$ & $+0.396$ $[+0.360,\,+0.429]^{*}$ \\
 & DSM & $+0.125$ $[+0.108,\,+0.142]^{*}$ & $+0.509$ $[+0.484,\,+0.536]^{*}$ \\
 & DCPH & $+0.053$ $[+0.039,\,+0.068]^{*}$ & $+0.374$ $[+0.345,\,+0.404]^{*}$ \\
 & survTLS & $+0.044$ $[+0.034,\,+0.054]^{*}$ & $+0.431$ $[+0.410,\,+0.453]^{*}$ \\
 & RF & $+0.068$ $[+0.053,\,+0.083]^{*}$ & $+0.495$ $[+0.461,\,+0.532]^{*}$ \\
 & Hard EEP ($H=h_{\min}$) & $+0.036$ $[+0.028,\,+0.044]^{*}$ & $+0.302$ $[+0.283,\,+0.320]^{*}$ \\
 & Hard EEP ($H=h_{\max}$) & $+0.028$ $[+0.021,\,+0.035]^{*}$ & $+0.157$ $[+0.136,\,+0.177]^{*}$ \\
\bottomrule
\end{tabular}

%% file: manuscript/tables/calibration_summary.tex
\begin{tabular}{llrr}
\toprule
Device & Method & \makecell{Median\\test F1} & \makecell{95\% interval\\of test F1} \\
\midrule
\multirow{7}{*}{C-Mod} & CCNN & $0.162$ & $[0.051,\,0.191]$ \\
 & DSM & $0.135$ & $[0.022,\,0.249]$ \\
 & DCPH & $0.157$ & $[0.093,\,0.268]$ \\
 & survTLS & $0.113$ & $[0.084,\,0.210]$ \\
 & RF & $0.293$ & $[0.164,\,0.359]$ \\
 & Hard EEP & $0.207$ & $[0.059,\,0.309]$ \\
 & TLS & $0.265$ & $[0.055,\,0.366]$ \\
\midrule
\multirow{7}{*}{DIII-D} & CCNN & $0.129$ & $[0.124,\,0.159]$ \\
 & DSM & $0.236$ & $[0.169,\,0.316]$ \\
 & DCPH & $0.184$ & $[0.115,\,0.313]$ \\
 & survTLS & $0.216$ & $[0.192,\,0.252]$ \\
 & RF & $0.316$ & $[0.267,\,0.336]$ \\
 & Hard EEP & $0.113$ & $[0.079,\,0.258]$ \\
 & TLS & $0.420$ & $[0.288,\,0.526]$ \\
\midrule
\multirow{7}{*}{EAST} & CCNN & $0.387$ & $[0.373,\,0.419]$ \\
 & DSM & $0.282$ & $[0.092,\,0.375]$ \\
 & DCPH & $0.415$ & $[0.255,\,0.542]$ \\
 & survTLS & $0.377$ & $[0.343,\,0.551]$ \\
 & RF & $0.286$ & $[0.115,\,0.286]$ \\
 & Hard EEP & $0.493$ & $[0.377,\,0.660]$ \\
 & TLS & $0.780$ & $[0.614,\,0.826]$ \\
\bottomrule
\end{tabular}

%% file: manuscript/tables/multi_deadline.tex
\begin{tabular}{llrrrr}
\toprule
Device & Method & $w=10$ & $w=40$ & $w=100$ & $w=300$ \\
\midrule
\multirow{7}{*}{C-Mod} & CCNN & 0.436 & 0.430 & 0.394 & 0.281 \\
 & DSM & 0.686 & 0.581 & 0.446 & 0.187 \\
 & DCPH & 0.635 & 0.562 & 0.472 & 0.230 \\
 & survTLS & 0.768 & 0.600 & 0.470 & 0.228 \\
 & RF & 0.804 & 0.722 & 0.642 & 0.270 \\
 & Hard EEP & 0.744 & 0.691 & 0.620 & 0.228 \\
 & TLS & 0.745 & 0.696 & 0.626 & 0.279 \\
\midrule
\multirow{7}{*}{DIII-D} & CCNN & 0.314 & 0.302 & 0.276 & 0.220 \\
 & DSM & 0.812 & 0.731 & 0.611 & 0.436 \\
 & DCPH & 0.795 & 0.725 & 0.631 & 0.504 \\
 & survTLS & 0.718 & 0.642 & 0.556 & 0.404 \\
 & RF & 0.886 & 0.807 & 0.648 & 0.377 \\
 & Hard EEP & 0.871 & 0.686 & 0.448 & 0.300 \\
 & TLS & 0.927 & 0.819 & 0.495 & 0.315 \\
\midrule
\multirow{7}{*}{EAST} & CCNN & 0.616 & 0.613 & 0.606 & 0.584 \\
 & DSM & 0.894 & 0.850 & 0.743 & 0.559 \\
 & DCPH & 0.943 & 0.922 & 0.871 & 0.770 \\
 & survTLS & 0.961 & 0.931 & 0.852 & 0.479 \\
 & RF & 0.953 & 0.907 & 0.784 & 0.547 \\
 & Hard EEP & 0.963 & 0.939 & 0.835 & 0.432 \\
 & TLS & 0.984 & 0.975 & 0.934 & 0.403 \\
\bottomrule
\end{tabular}

%% file: manuscript/tables/protocol_variants.tex
\begin{tabular}{llcrr}
\toprule
Device & Variant & $H$, $M$ & AUPRC & F1 \\
\midrule
\multirow{4}{*}{C-Mod} & DSM, sampled & $40,\,1$ & 0.522 & 0.145 \\
 & DSM, all (main) & $80,\,3$ & 0.581 & 0.139 \\
 & survTLS, original & $80,\,\text{--}$ & 0.611 & 0.141 \\
 & survTLS, resel. (main) & $80,\,\text{--}$ & 0.600 & 0.118 \\
\midrule
\multirow{4}{*}{DIII-D} & DSM, sampled & $20,\,1$ & 0.648 & 0.203 \\
 & DSM, all (main) & $120,\,1$ & 0.731 & 0.220 \\
 & survTLS, original & $120,\,\text{--}$ & 0.653 & 0.180 \\
 & survTLS, resel. (main) & $40,\,\text{--}$ & 0.642 & 0.189 \\
\midrule
\multirow{4}{*}{EAST} & DSM, sampled & $80,\,3$ & 0.817 & 0.144 \\
 & DSM, all (main) & $80,\,3$ & 0.850 & 0.272 \\
 & survTLS, original & $20,\,\text{--}$ & 0.761 & 0.080 \\
 & survTLS, resel. (main) & $360,\,\text{--}$ & 0.931 & 0.350 \\
\bottomrule
\end{tabular}

%% file: manuscript/tables/disruptionbench_metrics.tex
\begin{tabular}{llrrrr}
\toprule
Device & Method & AUC (Zhu) & TPR & FPR & F1 \\
\midrule
\multirow{7}{*}{C-Mod} & CCNN & 0.722 & 0.094 & 0.014 & 0.164 \\
 & DSM & 0.808 & 0.098 & 0.004 & 0.172 \\
 & DCPH & 0.803 & 0.127 & 0.010 & 0.205 \\
 & survTLS & 0.817 & 0.103 & 0.006 & 0.179 \\
 & RF & 0.884 & 0.293 & 0.014 & 0.435 \\
 & Hard EEP & 0.866 & 0.138 & 0.003 & 0.234 \\
 & TLS & 0.871 & 0.228 & 0.008 & 0.362 \\
\midrule
\multirow{7}{*}{DIII-D} & CCNN & 0.702 & 0.081 & 0.015 & 0.131 \\
 & DSM & 0.904 & 0.188 & 0.002 & 0.307 \\
 & DCPH & 0.899 & 0.136 & 0.003 & 0.232 \\
 & survTLS & 0.849 & 0.138 & 0.000 & 0.242 \\
 & RF & 0.933 & 0.274 & 0.002 & 0.423 \\
 & Hard EEP & 0.902 & 0.039 & 0.001 & 0.074 \\
 & TLS & 0.935 & 0.409 & 0.007 & 0.516 \\
\midrule
\multirow{7}{*}{EAST} & CCNN & 0.698 & 0.290 & 0.041 & 0.402 \\
 & DSM & 0.915 & 0.222 & 0.002 & 0.339 \\
 & DCPH & 0.963 & 0.331 & 0.003 & 0.489 \\
 & survTLS & 0.964 & 0.340 & 0.001 & 0.367 \\
 & RF & 0.955 & 0.217 & 0.000 & 0.357 \\
 & Hard EEP & 0.963 & 0.437 & 0.001 & 0.543 \\
 & TLS & 0.984 & 0.756 & 0.003 & 0.848 \\
\bottomrule
\end{tabular}